\documentclass[a4paper,11pt]{article}
\usepackage{jheppub}

\usepackage{amssymb,amsmath}
\usepackage{enumitem}
\usepackage[normalem]{ulem}
\usepackage{comment}
\usepackage{float}
\usepackage{braket}

\usepackage[subrefformat=parens]{subcaption}
\newcommand{\eV}{\mbox{eV}}
\newcommand{\MeV}{\mbox{MeV}}

\newcommand{\GeV}{\mbox{GeV}}
\newcommand{\SU}{{\rm SU}}

\newcommand{\2}{\mathbf{2}}
\newcommand{\Umt}{\mathrm{U(1)}_{L_\mu-L_\tau}}
\newcommand{\CP}{{\rm CP}}

\def\lsim{\mathrel{\rlap{\lower4pt\hbox{\hskip1pt$\sim$}}
    \raise1pt\hbox{$<$}}}         %less than or approx. symbol
\def\gsim{\mathrel{\rlap{\lower4pt\hbox{\hskip1pt$\sim$}}
    \raise1pt\hbox{$>$}}}         %greater than or approx. symbol

\numberwithin{equation}{section}

\preprint{
\begin{minipage}{10cm}
\small
\flushright
KYUSHU-HET-280, 
KEK-TH-2597
\end{minipage}}

\title{New Constraints on Gauged U(1)$_{L_\mu-L_\tau}$ Models via $Z-Z'$ Mixing}

\author[1]{Kento Asai,} 
\author[2]{Coh Miyao,} 
\author[3,4]{Shohei Okawa,} 
\author[2]{Koji Tsumura} 
\affiliation[1]{Institute for Cosmic Ray Research (ICRR), The University of Tokyo,\\ Kashiwa, Chiba 277-8582, Japan}
\affiliation[2]{Department of Physics, Kyushu University, 744 Motooka, Nishi-ku, Fukuoka 819-0395, Japan}
\affiliation[3]{KEK Theory Center, Tsukuba, Ibaraki 305-0801, Japan}
\affiliation[4]{Departament de F\'isica Qu\`antica i Astrof\'isica, Institut de Ci\`encies del Cosmos (ICCUB),\\
Universitat de Barcelona, Mart\'i i Franqu\`es 1, E-08028 Barcelona, Spain\\}

\emailAdd{kento@icrr.u-tokyo.ac.jp}
\emailAdd{miyao.coh@phys.kyushu-u.ac.jp}
\emailAdd{okawash@post.kek.jp}
\emailAdd{tsumura.koji@phys.kyushu-u.ac.jp}

\abstract{
Models based on a U(1)$_{L_\mu-L_\tau}$ gauge symmetry can explain the discrepancy between 
the measured value and theoretical prediction of the muon anomalous magnetic moment. 
Based on the latest experimental results, we revisit the analysis of neutrino mass matrix structures 
in minimal U(1)$_{L_\mu-L_\tau}$ models where the U(1)$_{L_\mu-L_\tau}$ symmetry breaking is caused by a single scalar field. 
Due to the charge assignment of the scalar field, each model predicts a unique structure of the neutrino mass matrix, 
which demands non-trivial relations among neutrino mass and mixing parameters. 
We find that the model called type ${\bf 2}_{+1}$, which features an SU(2)$_L$ doublet scalar $\Phi_{+1}$ with the U(1)$_{L_\mu - L_\tau}$ charge $+1$ and the hypercharge $+1/2$ and predicts the $\bf B_3$ texture structure, is marginally acceptable under the current neutrino oscillation data and cosmological observation.
When the U(1)$_{L_\mu - L_\tau}$ gauge symmetry is broken by the vacuum expectation value of the standard model 
non-singlet representation such as $\Phi_{+1}$, there are additional contributions to flavor-changing meson decay processes 
and atomic parity violation via mixing between the $Z$ boson and the U(1)$_{L_\mu-L_\tau}$ gauge boson $Z'$. 
We newly evaluate the model-dependent constraints on the model and 
find that the type ${\bf 2}_{+1}$ model is robustly ruled out. 
The model is extended to have an additional vacuum expectation value of a standard model singlet scalar 
in order to avoid the stringent constraints via the $Z-Z'$ mixing. 
Finally, we clarify the allowed range of the ratio of these vacuum expectation values.
}

\makeatletter
\gdef\@fpheader{}
\makeatother

\begin{document}

\maketitle

%%%%%%%%%%%%%%%%%%%%%%%%%%%%%%%%%%%%%%%%%%%%%%%%%%%%%%%%%%%%%%%%%%%%%%%%%%%%%%%%%%%%%%%%%%%%%%%%%%%%%%%%%%
\section{Introduction}

The discovery of the Higgs boson in 2012~\cite{ATLAS:2012yve,CMS:2012qbp} has established the Standard Model (SM) of particle physics. It has successfully explained a variety of experimental results so far. 
Nonetheless, the SM is not considered a theory of everything due to theoretical issues and empirical problems, 
that indicates the extension of the SM. 

In recent years, several experimental hints for physics beyond the SM have come out. 
Among them is a discrepancy in the muon anomalous magnetic moment, also known as the muon $g-2$ anomaly. 
The world average value of the measured muon $g-2$ at Brookhaven National Laboratory~\cite{Muong-2:2002wip,Muong-2:2004fok,Muong-2:2006rrc} and Fermilab~\cite{Muong-2:2021ojo,Muong-2:2023cdq} disagrees with the SM prediction. 
The prediction crucially depends on the evaluation of the hadronic vacuum polarization (HVP) contribution to the muon $g-2$. 
Together with the HVP contribution based on the dispersive analysis using $e^+e^- \to \mbox{hadrons}$ cross-section data until 2020, a $4.3\sigma$ discrepancy has been known~\cite{Aoyama:2020ynm}, 
\begin{align}
    \Delta a_\mu = a_\mu^{\rm exp} - a_\mu^{\rm SM} = (25.1 \pm 5.9) \times 10^{-10}.
\end{align}
On the other hand, the BMW collaboration recently reported an update on lattice QCD calculations of the leading order HVP contribution~\cite{Borsanyi:2020mff}. 
They obtained a value closer to the experimental average, which mitigates the discrepancy up to the $\sim2\sigma$ level. 
Their result is partially supported by other lattice calculations afterwards~\cite{Ce:2022kxy, ExtendedTwistedMass:2022jpw}. 
In the dispersive approach, new data sets in the $e^+e^- \to \mbox{hadrons}$ scattering are available after ref.~\cite{Aoyama:2020ynm}. 
The CMD-3 collaboration reported a new measurement of $e^+e^- \to \pi^+ \pi^-$ cross-section~\cite{CMD-3:2023alj, CMD-3:2023rfe}. 
They obtained a similar value to the BMW result, yet this new CMD-3 result disagrees at the 2.5\,--\,5$\sigma$ level with the previous dispersive determinations. 
The BABAR~\cite{BABAR:2021cde} and Belle-II~\cite{Belle-II:2024msd} collaborations separately reported measurements of $e^+e^- \to \pi^+ \pi^- \pi^0$ cross-section. 
Their results disagree with each other at the $2.5\sigma$ level. 
The origin of these disagreements in the dispersive approach is still unknown. 
Therefore, while there is still plenty of room for SM explanations, we need more precise cross-section data and ongoing efforts by lattice groups to clarify the situation.

If the muon $g-2$ discrepancy is taken as a clue to physics beyond the SM, 
it is %thus 
natural to ask what new physics models can accommodate this discrepancy. 
A model with a gauged U(1)$_{L_\mu - L_\tau}$ symmetry is one such candidate~\cite{Foot:1990mn,He:1990pn,He:1991qd,Foot:1994vd}.
In this model, the U(1)$_{L_\mu - L_\tau}$ gauge boson $Z'$ interacts with muon, but not with electron and quarks at the tree level. 
The correction to the muon $g-2$ from this new gauge interaction can close the gap in the value of the muon $g-2$  
without conflicting other experimental bounds~\cite{Baek:2001kca,Ma:2001md,Heeck:2011wj,Harigaya:2013twa}, although the NA64$\mu$ experiment has recently reported a strong bound on the $Z'$ boson~\cite{Andreev:2024sgn}.

Neutrino mass is another problem in the SM. 
In order to fit the measured pattern of the neutrino oscillation parameters, 
the neutrino sector in the SM has to be extended such that at least two neutrinos have non-zero masses. 
One of the simplest neutrino mass generation mechanisms is the so-called seesaw mechanism~\cite{Minkowski:1977sc,Yanagida:1979as,Gell-Mann:1979vob,Mohapatra:1979ia}, where heavy Majorana right-handed neutrinos are added to the SM. 

In models based on the U(1)$_{L_\mu - L_\tau}$ gauge symmetry, 
the neutrino mass generation is non-trivially realized  
because the U(1)$_{L_\mu - L_\tau}$ symmetry restricts neutrino mass matrix structures. 
In fact, the addition of only three right-handed neutrinos is not enough to explain the measured neutrino oscillation data. 
To solve this problem in an economical way, one can newly introduce a U(1)$_{L_\mu - L_\tau}$ charged scalar that develops a non-zero vacuum expectation value (VEV). 
In refs.~\cite{Asai:2017ryy,Asai:2018ocx,Asai:2019ciz}, 
the following three {\it minimal} models are considered in this direction:
\begin{itemize}
    \item Type $\mathbf{1}$ : an SU(2)$_L$ singlet scalar $\sigma$ with the U(1)$_{L_\mu - L_\tau}$ charge $+1$,
    \item Type $\mathbf{2}_{+1}$ : an SU(2)$_L$ doublet scalar $\Phi_{+1}$ with the U(1)$_{L_\mu - L_\tau}$ charge $+1$,
    \item Type $\mathbf{2}_{-1}$ : an SU(2)$_L$ doublet scalar $\Phi_{-1}$ with the U(1)$_{L_\mu - L_\tau}$ charge $-1$.
\end{itemize}
The hypercharge of the additional singlet (doublet) scalar field $\sigma$ $(\Phi_{\pm 1})$ is taken to be $0$ $(+1/2)$. 
In these minimal models, 
the mass matrix for the light neutrinos has a two-zero texture (minor) structure, that is, the (inverse) mass matrix for the light neutrinos has two independent vanishing elements. 
This unique two-zero structure leads to close relations among the neutrino oscillation parameters. 
Using these relations, 
undetermined parameters, i.e., 
the neutrino Dirac and Majorana CP phases and the sum of the neutrino masses, are predicted from the well-measured neutrino mixing angles and mass squared differences. 
In refs.~\cite{Asai:2017ryy,Asai:2018ocx,Asai:2019ciz}, 
it is shown that in the type $\mathbf{1}$ case, 
the sum of the neutrino masses predicted by using the \texttt{NuFITv4.0} global analysis~\cite{Esteban:2018azc} is consistent with the Planck 2018 limit~\cite{Planck:2018vyg} at the $3\sigma$ level for the normal mass ordering.\footnote{
For the inverted mass ordering, the type $\mathbf{1}$ case cannot reproduce the observed neutrino parameters.} 
On the other hand, the type $\mathbf{2}_{+1}$ and type $\mathbf{2}_{-1}$ cases are excluded by the Planck 2018 limit, regardless of the neutrino mass orderings. 

In the meantime, new neutrino oscillation data have become available. 
Accordingly, 
the predictions for the neutrino oscillation parameters in the minimal gauged U(1)$_{L_\mu - L_\tau}$ models will be updated. 
In this paper, therefore, we revisit these minimal models based on the results of the latest \texttt{NuFITv5.2} global analysis~\cite{nufit}. 
Then we find that the type $\mathbf{2}_{+1}$ case revives with the latest neutrino global parameters. 

Besides updating the neutrino analysis, we carefully study new constraints in the minimal models. 
Since the newly introduced SU(2)$_L$ doublet scalar $\Phi_{+1}$ breaks not only the U(1)$_{L_\mu-L_\tau}$ gauge symmetry 
but also the electroweak (EW) symmetry, the $Z'$ boson mixes with the $Z$ boson in their mass term~\cite{Babu:1997st}.
After diagonalizing the mass matrix, the $Z'$ boson couples to the SM neutral current, 
resulting in a new contribution to atomic parity violation (APV)~\cite{Davoudiasl:2012ag,Davoudiasl:2012qa,Davoudiasl:2014kua}. 
Moreover, the coupling to the neutral current also induces 
flavor-changing meson decay processes~\cite{Davoudiasl:2012ag,Davoudiasl:2014kua}. 
In this paper, we analyze the new constraints on the gauged U$(1)_{L_\mu - L_\tau}$ models 
from these phenomena via the $Z - Z'$ mixing.

This paper is organized as follows. 
In section~\ref{sec:model}, we first introduce the minimal gauged U(1)$_{L_\mu - L_\tau}$ models. 
We reanalyze the neutrino mass matrix structures in section~\ref{sec:analysis}. 
In section~\ref{sec:constraint}, we consider the new model-dependent constraints on the gauged U(1)$_{L_\mu - L_\tau}$ models with the additional SU(2)$_L$ doublet scalar. 
Finally, we summarize our results in section~\ref{sec:conclusion}.

%%%%%%%%%%%%%%%%%%%%%%%%%%%%%%%%%%%%%%%%%%%%%%%%%%%%%%%%%%%%%%%%%%%%%%%%%%%%%%%%%%%%%%%%%%%%%%%%%%%%%%%%%%
\section{Types of minimal models}
\label{sec:model}

In this section, we introduce the minimal gauged U(1)$_{L_\mu - L_\tau}$ models~\cite{Asai:2017ryy,Asai:2018ocx}, where three right-handed neutrinos  $N_i\ (i=e,\mu,\tau)$ and a single U(1)$_{L_\mu - L_\tau}$-breaking scalar are added to the SM. 
The minimal models are classified according to the SU(2)$_L$ representation and the U(1)$_{L_\mu - L_\tau}$ charge of the additional scalar. 
In this section, we summarize the matter content, relevant interactions, and characteristics of neutrino mass matrix structures in the type $\mathbf{2}_{+1}$ and $\mathbf{2}_{-1}$ models. 
Details for the type $\mathbf{1}$ model are provided in appendix~\ref{app:Cminor}.

\subsection{\texorpdfstring{Type $\mathbf{2}_{+1}$}{Type 2\_+1}}

In addition to the SM fields and three right-handed neutrinos $N_i$, 
we introduce an SU$(2)_L$ doublet scalar field $\Phi_{+1}$ with the U$(1)_{L_\mu - L_\tau}$ charge $+1$ and the hypercharge $+1/2$ in the type $\mathbf{2}_{+1}$ model. 
In table~\ref{tab:model2-1_contents}, we list the U(1)$_{L_\mu - L_\tau}$ charges and SU(2)$_L$ representations 
of the lepton and scalar fields. 
%%%%%%%%%%%%%%%%%%% Table %%%%%%%%%%%%%%%%%%%
\begin{table}[t]
    \centering
    \begin{tabular}{|c||c|c|c|c|c|}\hline
         Fields & $(L_e, L_\mu, L_\tau)$ & $(e_R^{}, \mu_R^{}, \tau_R^{})$ & $(N_e, N_\mu, N_\tau)$ & $H$ & $\Phi_{+1}$ \\ \hline
         U$(1)_{L_\mu - L_\tau}$ & $(0,+1,-1)$ & $(0,+1,-1)$ & $(0,+1,-1)$ & $0$ & $+1$ \\ \hline
          SU$(2)_L$ & $\bf 2$ & $\bf 1$ & $\bf 1$ & $\bf 2$ & $\bf 2$ \\ \hline
    \end{tabular}
    \caption{Charge assignments of the relevant fields in the type $\mathbf{2}_{+1}$ model.}
    \label{tab:model2-1_contents}
\end{table}
%%%%%%%%%%%%%%%%%%%%%%%%%%%%%%%%%%%%%%%%%%%%%
The Lagrangian for the leptonic Yukawa interactions and mass terms in this model is written by
\begin{align}
\label{eq:Lag-typeII1}
    \mathcal{L} \supset& -y_e\, e^c_R (L_e \cdot H^\dagger) -y_\mu\, \mu^c_R (L_\mu \cdot H^\dagger) -y_\tau\, \tau^c_R (L_\tau \cdot H^\dagger) \nonumber \\
    &-y_{\mu e}\, e^c_R (L_\mu \cdot \Phi_{+1}^\dagger) -y_{e\tau}\, \tau^c_R (L_e \cdot \Phi_{+1}^\dagger) \nonumber \\ 
    &-\lambda_e\, N_e^c (L_e \cdot H) -\lambda_\mu\, N_\mu^c (L_\mu \cdot H) -\lambda_\tau\, N_\tau^c (L_\tau \cdot H) \nonumber \\
    &-\lambda_{\tau e}\, N^c_e (L_\tau\cdot \Phi_{+1}) - \lambda_{e \mu}\, N^c_\mu (L_e\cdot \Phi_{+1}) \nonumber \\
    &-\frac{1}{2} M_{ee} N^c_e N^c_e -M_{\mu \tau}N^c_\mu N^c_\tau +{\rm H.c.},
\end{align}
where $L_i\ (i=e,\mu,\tau)$ are the left-handed lepton doublets, $(e_R^{},\mu_R^{},\tau_R^{})$ are the right-handed lepton singlets, and $H$ is the Higgs doublet in the SM. 
In eq.~\eqref{eq:Lag-typeII1}, the middle dots ($\cdot$) between the SU(2)$_L$ doublet fields indicate the contraction of the SU(2)$_L$ indices.
It is assumed that $H$ and $\Phi_{+1}$ acquire non-zero VEVs as 
\begin{align}
\braket{H} = \frac{1}{\sqrt{2}} \begin{pmatrix}
    0 \\
    v_1
\end{pmatrix},\ \ \ \ 
\braket{\Phi_{+1}} = \frac{1}{\sqrt{2}} \begin{pmatrix}
    0 \\
    v_2
\end{pmatrix},
\label{eq:vev-2+}
\end{align}
where $\sqrt{v_1^2 + v_2^2} = v \simeq 246\,\GeV$ is satisfied. 
After the EW and U$(1)_{L_\mu - L_\tau}$ symmetry breaking, 
the above Yukawa interaction terms lead to the neutrino Dirac and Majorana mass terms:
\begin{align}
    \mathcal{L} \supset & 
    -
    \begin{pmatrix} \nu_e & \nu_\mu & \nu_\tau \end{pmatrix} 
    \mathcal{M}_{D}
    \begin{pmatrix} N_e^c \\ N_\mu^c \\ N_\tau^c \end{pmatrix} 
    -\frac{1}{2}
    \begin{pmatrix} N_e^c & N_\mu^c & N_\tau^c \end{pmatrix} 
    \mathcal{M}_{R}
    \begin{pmatrix} N_e^c \\ N_\mu^c \\ N_\tau^c \end{pmatrix} 
    + {\rm H.c.},
\end{align}
where 
\begin{align}
    \mathcal{M}_{D}
    &=\frac{1}{\sqrt{2}}
    \begin{pmatrix}
        \lambda_e v_1 & \lambda_{e \mu} v_2 & 0 \\
        0 & \lambda_\mu v_1 & 0 \\
        \lambda_{\tau e} v_2 & 0 & \lambda_{\tau} v_1
    \end{pmatrix},
    \quad
    \mathcal{M}_{R}
    =
    \begin{pmatrix}
        M_{ee} & 0 & 0 \\
        0 & 0 & M_{\mu \tau} \\
        0 & M_{\mu \tau} & 0
    \end{pmatrix}.
\end{align}
Then, the mass matrix for the light neutrinos is given by the seesaw mechanism~\cite{Minkowski:1977sc,Yanagida:1979as,Gell-Mann:1979vob,Mohapatra:1979ia} as 
\begin{align}
    \mathcal{M}_{\nu} &\simeq - \mathcal{M}_{D} \mathcal{M}_{R}^{-1} \mathcal{M}_{D}^T \nonumber \\
    &=
    \begin{pmatrix}
         -\frac{\lambda_e^2 v_1^2}{2 M_{ee}} & 0 & - \frac{v_1 v_2 \left(\lambda_e \lambda_{\tau e} M_{\mu \tau} +  \lambda_{\tau} \lambda_{e\mu} M_{ee} \right)}{2 M_{ee} M_{\mu \tau}} \\
         0 & 0 & -\frac{\lambda_{\mu} \lambda_{\tau} v_1^2}{2 M_{\mu \tau}} \\
         -\frac{v_1 v_2 \left(\lambda_e \lambda _{\tau e} M_{\mu \tau} + \lambda_{\tau} \lambda_{e\mu} M_{ee} \right)}{2 M_{ee} M_{\mu \tau}} & -\frac{\lambda_{\mu} \lambda_{\tau} v_1^2}{2 M_{\mu \tau}} & -\frac{\lambda_{\tau e}^2 v_2^2}{2 M_{ee}} 
    \end{pmatrix}.
\label{eq:Mnu-B3}
\end{align}
The mass matrix is symmetric and has two independent zeros in the $(1,2)$ and $(2,2)$ elements. 
This structure is called the ${\bf B}_3$ texture~\cite{Frampton:2002yf,Xing:2002ta,Xing:2002ap,Guo:2002ei,Fritzsch:2011qv,Araki:2012ip}. 
We note that the Yukawa coupling matrix for the charged leptons is assumed to be almost diagonal in order to avoid strong constraints from flavor-violating decays of tau leptons, see refs.~\cite{Asai:2017ryy,Asai:2018ocx} for the details.

\subsection{\texorpdfstring{Type $\mathbf{2}_{-1}$}{Type 2\_-1}}

The type $\mathbf{2}_{-1}$ model contains an SU(2)$_L$ doublet scalar $\Phi_{-1}$ with the U(1)$_{L_\mu - L_\tau}$ charge $-1$ and the hypercharge $+1/2$.
The U(1)$_{L_\mu - L_\tau}$ charges and SU(2)$_L$ representations of the lepton and scalar fields are shown in table~\ref{tab:model2-2_contents}.
%%%%%%%%%%%%%%%%%%% Table %%%%%%%%%%%%%%%%%%%
\begin{table}[t]
    \centering
    \begin{tabular}{|c||c|c|c|c|c|}\hline
         Fields & $(L_e, L_\mu, L_\tau)$ & $(e_R^{}, \mu_R^{}, \tau_R^{})$ & $(N_e, N_\mu, N_\tau)$ & $H$ & $\Phi_{-1}$ \\ \hline
         U(1)$_{L_\mu - L_\tau}$ & $(0,+1,-1)$ & $(0,+1,-1)$ & $(0,+1,-1)$ & $0$ & $-1$ \\ \hline
          SU(2)$_L$ & $\bf 2$ & $\bf 1$ & $\bf 1$ & $\bf 2$ & $\bf 2$ \\ \hline
    \end{tabular}
    \caption{Charge assignments of the relevant fields in the type $\mathbf{2}_{-1}$ model.}
    \label{tab:model2-2_contents}
\end{table}
%%%%%%%%%%%%%%%%%%%%%%%%%%%%%%%%%%%%%%%%%%%%%
The Lagrangian for the leptonic Yukawa interactions and the mass terms is given by
\begin{align}
    \mathcal{L} \supset& 
    -y_e\, e^c_R (L_e \cdot H^\dagger) -y_\mu\, \mu^c_R (L_\mu \cdot H^\dagger) -y_\tau\, \tau^c_R (L_\tau \cdot H^\dagger)
    \nonumber \\
    &-y_{\tau e}\, e^c_R (L_\tau \cdot \Phi_{-1}^\dagger) -y_{e\mu}\, \mu^c_R (L_e \cdot \Phi_{-1}^\dagger) \nonumber \\ 
    &-\lambda_e\, N_e^c (L_e \cdot H) -\lambda_\mu\, N_\mu^c (L_\mu \cdot H) -\lambda_\tau\, N_\tau^c (L_\tau \cdot H) \nonumber \\
    &- \lambda_{\mu e}\, N^c_e (L_\mu\cdot \Phi_{-1}) - \lambda_{e \tau}\, N^c_\tau (L_e\cdot \Phi_{-1}) \nonumber \\
    &-\frac{1}{2} M_{ee} N^c_e N^c_e -M_{\mu \tau}N^c_\mu N^c_\tau +{\rm H.c.}
\end{align}
The U$(1)_{L_\mu - L_\tau}$-breaking doublet scalar $\Phi_{-1}$ acquires a non-zero VEV:
\begin{align}
\braket{\Phi_{-1}} = \frac{1}{\sqrt{2}} \begin{pmatrix}
    0 \\
    v_2
\end{pmatrix}.
\end{align}
We use the same symbol $v_2$ for the VEV of $\Phi_{-1}$ as that of $\Phi_{+1}$ because this does not cause any confusion in the following analysis.
After $H$ and $\Phi_{-1}$ develop the VEVs, the Dirac neutrino mass terms emerge from the Yukawa interactions. 
The mass matrix for the light neutrinos is then given by 
\begin{align}
    \mathcal{M}_{\nu} &\simeq - \mathcal{M}_{D} \mathcal{M}_{R}^{-1} \mathcal{M}_{D}^T \nonumber \\
    &=
    \begin{pmatrix}
    -\frac{\lambda _e^2 v_1^2}{2 M_{ee}} & -\frac{v_1 v_2 \left(\lambda_e \lambda_{\mu e} M_{\mu \tau} + M_{ee} \lambda_{\mu} \lambda_{e\tau} \right)}{2 M_{ee} M_{\mu \tau}} & 0 \\
    -\frac{v_1 v_2 \left(\lambda_e \lambda_{\mu e} M_{\mu \tau} + \lambda_{e\tau} \lambda_{\mu} M_{ee} \right)}{2 M_{ee} M_{\mu \tau}} & -\frac{\lambda_{\mu e}^2 v_2^2}{2 M_{ee}} & -\frac{\lambda_{\mu} \lambda_{\tau} v_1^2}{2 M_{\mu \tau}} \\
    0 & -\frac{\lambda_{\mu} \lambda_{\tau} v_1^2}{2 M_{\mu \tau}} & 0 \\
    \end{pmatrix},
\label{eq:Mnu-B4}
\end{align}
where 
\begin{align}
    \mathcal{M}_{D}
    &=\frac{1}{\sqrt{2}}
    \begin{pmatrix}
        \lambda_e v_1 & 0 & \lambda_{e \tau} v_2 \\
        \lambda_{\mu e} v_2 & \lambda_\mu v_1 & 0 \\
        0 & 0 & \lambda_{\tau} v_1
    \end{pmatrix},
    \quad
    \mathcal{M}_{R}
    =
    \begin{pmatrix}
        M_{ee} & 0 & 0 \\
        0 & 0 & M_{\mu \tau} \\
        0 & M_{\mu \tau} & 0
    \end{pmatrix}.
\end{align}
In the type $\2_{-1}$ case, 
the mass matrix for the light neutrinos has zeros in the $(1,3)$ and $(3,3)$ elements. 
This structure is known as the ${\bf B}_4$ texture.
The off-diagonal elements of the charged lepton mass matrix are assumed to be negligible as in the type $\mathbf{2_{+1}}$ case.

\subsection{\texorpdfstring{Correction to Muon $g-2$}{Correction to Muon g-2}}

The U$(1)_{L_\mu - L_\tau}$ gauge boson $Z^\prime$ gives a sizable correction to the muon $g-2$. 
Notably, the muon $g-2$ discrepancy between the SM prediction and experimental value can be resolved when the $Z'$ mass is $m_{Z'}\sim 10 \, \text{--} \, 40\,\MeV$~\cite{Baek:2001kca,Ma:2001md,Heeck:2011wj,Harigaya:2013twa,Andreev:2024sgn}. 
The one-loop contribution to the muon $g-2$ is given by 
\begin{align}
    \Delta a_\mu = \frac{g_{Z'}^2}{8\pi^2} \int_0^1\ {\rm d}x \frac{2 x^2 (1-x) m_\mu^2}{x^2 m_\mu^2 +(1-x)m_{Z'}^2},
\end{align}
where $g_{Z'}^{}$ is the gauge coupling constant of the U(1)$_{L_\mu - L_\tau}$ gauge symmetry.

%%%%%%%%%%%%%%%%%%%%%%%%%%%%%%%%%%%%%%%%%%%%%%%%%%%%%%%%%%%%%%%%%%%%%%%%%%%%%%%%%%%%%%%%%%%%%%%%%%%%%%%%%%
\section{Revisiting Analysis for Two-Zero Neutrino Mass Matrix Structures}
\label{sec:analysis}

Neutrino mass matrix or its inverse matrix often has non-trivial structures in new physics models with a lepton flavor symmetry. 
As seen in section \ref{sec:model}, the neutrino mass matrices in the minimal U(1)$_{L_\mu-L_\tau}$ models have two zero elements except for their symmetric counterparts. 
Such matrix structures are known to provide a prediction for undetermined neutrino oscillation parameters, i.e., the neutrino Dirac and Majorana CP phases and the sum of the neutrino masses in terms of known neutrino oscillation parameters. 
In the minimal U(1)$_{L_\mu-L_\tau}$ models, 
three neutrino mass matrix structures, called the ${\bf C}$ minor, ${\bf B}_3$ texture, and ${\bf B}_4$ texture, are obtained~\cite{Asai:2018ocx}:
\begin{equation}
{\bf C}: 
\begin{pmatrix}
    * & * & * \\
    * & 0 & * \\
    * & * & 0 
\end{pmatrix}
\,,\quad 
{\bf B}_3 : 
\begin{pmatrix}
    * & 0 & * \\
    0 & 0 & * \\
    * & * & * 
\end{pmatrix}
\,,\quad
{\bf B}_4: 
\begin{pmatrix}
    * & * & 0 \\
    * & * & * \\
    0 & * & 0 
\end{pmatrix}
\,.
\end{equation}
In this section, 
we outline how to extract the predictions for the neutrino oscillation parameters from the two-zero matrix structures. 
Then, we revisit the analysis for the ${\bf B}_3$ and ${\bf B}_4$ textures and update the predictions. 
See refs.~\cite{Asai:2017ryy,Asai:2018ocx} for further details of the analysis. 

\subsection{Methodology}
The mass matrix for the light neutrinos is diagonalized by a unitary matrix $U$ called the Pontecorvo--Maki--Nakagawa--Sakata (PMNS) matrix~\cite{Pontecorvo:1967fh,Pontecorvo:1957cp,Pontecorvo:1957qd,Maki:1962mu}: 
\begin{align}
\label{eq:two-zero-eqs}
    U^* {\rm diag}(m_1, m_2, m_3) U^\dag = \mathcal{M}_\nu,
\end{align}
where
\begin{equation}
    U \!=\! 
    \begin{pmatrix}
        c_{12} c_{13} & s_{12} c_{13} & s_{13} e^{-i\delta_\CP} \\
        -s_{12} c_{23} - c_{12} s_{13} s_{23}  e^{i\delta_\CP} & c_{12} c_{23} - s_{12} s_{13} s_{23}  e^{i\delta_\CP} & c_{13} s_{23}  \\
        s_{12} s_{23} - c_{12} s_{13} c_{23}  e^{i\delta_\CP} & -c_{12} s_{23} - s_{12} s_{13} c_{23}  e^{i\delta_\CP} & c_{13} c_{23} 
    \end{pmatrix}\!\!
    \begin{pmatrix}
        ~1~ & & \\
        & e^{i\alpha_2/2} & \\
        & & e^{i\alpha_3/2}
    \end{pmatrix},
\end{equation}
with $m_i$ being the mass eigenvalues of the light neutrinos, $c_{ij} \equiv \cos\theta_{ij}$ and $s_{ij} \equiv \sin\theta_{ij}$.
From this matrix relation eq.~\eqref{eq:two-zero-eqs}, 
the two-zero elements of $\mathcal{M}_\nu$ give two complex equations with respect to nine real neutrino parameters ($\theta_{12}$, $\theta_{13}$, $\theta_{23}$, $\delta_\CP$, $m_1$, $m_2$, $m_3$, $\alpha_2$, and $\alpha_3$). 
By solving these equations, four of the nine parameters are determined by the remaining five. 
Normally, the former four parameters are chosen to be less determined ones---$\delta_\CP$, $m_1$, and $\alpha_{2,3}$---and the latter five to be well measured ones---$\theta_{12}$, $\theta_{13}$, $\theta_{23}$, $\Delta m_{21}^2$, and $\Delta m_{3\ell}^2$, where $\Delta m_{3\ell}^2 \equiv m_3^2 -m^2_1$ for the normal ordering (NO) and $\Delta m_{3\ell}^2 \equiv m_3^2 -m^2_2$ for the inverted ordering (IO). 
Putting the measured values in the five parameters, we obtain sharp predictions for the four less-known parameters. 
It should be noted that the two complex equations follow from the two-zero elements of ${\cal M}_\nu$ and are independent of how non-zero elements of ${\cal M}_\nu$ are expressed in terms of UV model parameters. 

To extract the predictions, we take the following steps:
\begin{enumerate}
    \item Solve the two complex equations with respect to $e^{i \alpha_{2,3}}$. Then by taking their absolute values, two real equations are obtained in terms of the seven parameters, i.e., $\theta_{12}$, $\theta_{13}$, $\theta_{23}$, $\Delta m_{21}^2$, $\Delta m_{3\ell}^2$, $m_1$, and $\delta_\CP$. This step can be done analytically.
    
    \item Solve these two real equations with respect to $m_1$. Then we are left with one real equation in terms of $\delta_{\rm CP}$ and the other five parameters. 
    This step is also done analytically. 

    \item Numerically solve the remaining real equation with respect to $\delta_\CP$. 
    A prediction for $\delta_\CP$ is obtained for given input values of $\theta_{12}$, $\theta_{13}$, $\theta_{23}$, $\Delta m_{21}^2$ and $\Delta m_{3\ell}^2$. 
    The predictions for $m_1$ and $\alpha_{2,3}$ are in turn derived by putting the predicted value of $\delta_\CP$ back in the solutions for $m_1$ and $\alpha_{2,3}$ that we obtained beforehand. 
\end{enumerate}
In this way, ref.~\cite{Asai:2018ocx} systematically studied the predictions in the ${\bf C}$ minor, ${\bf B}_3$ texture and ${\bf B}_4$ texture cases, based on the \texttt{NuFITv4.0} global analysis of the neutrino oscillation parameters~\cite{Esteban:2018azc}. 
The obtained predictions were compared with the Planck limit on the sum of the light neutrino masses~\cite{Planck:2018vyg}. 
The ${\bf C}$ minor structure with the NO was consistent with the measured oscillation parameters and Planck limit. 
Meanwhile, it turned out that the ${\bf B}_3$ and ${\bf B}_4$ texture structures explained the neutrino parameters but were not consistent with the Planck limit at that time. 

%%%%%%%%%%%%%%%%%%% Table %%%%%%%%%%%%%%%%%%%
\begin{table}[tb]
\centering
\begin{tabular}{|c||cc|cc|} 
\hline
& 
\multicolumn{2}{|c|}{Normal Ordering} &
\multicolumn{2}{c|}{Inverted Ordering}
\\ 
& 
Best fit $\pm\ 1\sigma$& 
$3\sigma$ range&
Best fit $\pm$ $1\sigma$& 
$3\sigma$ range
\\ \hline \hline
$\theta_{12} [{}^\circ]$ & 
$33.41^{+0.75}_{-0.72}$ & $31.31\to35.74$&
$33.41^{+0.75}_{-0.72}$ & $31.31\to35.74$
\\ \hline
$\theta_{23} [{}^\circ]$ & 
$42.2^{+1.1}_{-0.9}$ & 
$39.7\to51.0$&
$49.0^{+1.0}_{-1.2}$ & 
$39.9\to51.5$
\\ \hline
$\theta_{13} [{}^\circ]$ & 
$8.58^{+0.11}_{-0.11}$ & 
$8.23\to8.91$&
$8.57^{+0.11}_{-0.11}$ & 
$8.23\to8.94$ 
\\ \hline
$\delta_{\rm CP} [{}^\circ]$ & 
$232^{+36}_{-26}$ & 
$144\to350$&
$276^{+22}_{-29}$ & 
$194\to344$ 
\\ \hline
$\frac{\Delta m_{21}^2}{10^{-5}{\rm eV}^2}$ & 
$7.41^{+0.21}_{-0.20}$ & 
$6.82\to8.03$&
$7.41^{+0.21}_{-0.20}$ & 
$6.82\to8.03$ 
\\ \hline
$\frac{\Delta m_{3\ell}^2}{10^{-3}{\rm eV}^2}$ & 
$+2.507^{+0.026}_{-0.027}$ & 
$+2.427\to+2.590$&
$-2.486^{+0.025}_{-0.028}$ & 
$-2.570\to-2.406$ 
\\ \hline
\end{tabular}
\caption{The global fit results for the neutrino oscillation parameters from \texttt{NuFITv5.2}~\cite{nufit}. 
The best fit point $\pm 1\sigma$ and $3\sigma$ ranges are tabulated for each neutrino mass ordering. Note that $\Delta m_{3\ell}^2 \equiv m_3^2 -m^2_1 > 0$ for the normal ordering and  $\Delta m_{3\ell}^2 \equiv m_3^2 -m^2_2 < 0$ for the inverted ordering.}
\label{tab:NuFIT5_2}
\end{table}
%%%%%%%%%%%%%%%%%%%%%%%%%%%%%%%%%%%%%%%%%%%%%

Since ref.~\cite{Asai:2018ocx}, a new global fit analysis, \texttt{NuFITv5.2}, of the neutrino oscillation parameters has been released~\cite{nufit}. 
Cosmological analysis has also been refined. 
The bounds on the sum of the light neutrino masses are evaluated under different assumptions of cosmological datasets and models, statistical treatments, and neutrino mass ordering~\cite{Planck:2018vyg,Vagnozzi:2017ovm,Capozzi:2017ipn,RoyChoudhury:2018gay,RoyChoudhury:2019hls,Ivanov:2019hqk,DES:2021wwk,Tanseri:2022zfe}.
Hence, we reanalyze these three mass matrix structures based on the latest \texttt{NuFITv5.2} data~\cite{nufit} (see also table~\ref{tab:NuFIT5_2}). 
The resulting predictions are compared with a new cosmological limit with the Planck 2018 TTTEEE+lowE+lensing+BAO datasets in the $\Lambda$CDM+$\sum m_\nu$ model in ref.~\cite{RoyChoudhury:2019hls}, which does not assume three light neutrinos are degenerate. 
In the following, we show the results for the ${\bf B}_3$ texture and ${\bf B}_4$ texture cases. 
The results for the ${\bf C}$ minor are provided in appendix~\ref{app:Cminor}.

%%%%%%%%%%%%%%%%%%%%%%%%%%%%%%%%%%%%%%%%%%%%%%%%%%%%%%%%%%%%%%%%%%%%%%%%%%%%%%%%%%%%%%%%%%%%%%%%%%%%%%%%%%
\subsection{\texorpdfstring{${\bf B}_3$ texture}{B3 texture}} 

%%%%%%%%%%%%%%%%%%% Figure %%%%%%%%%%%%%%%%%%%
\begin{figure}[tb]
    \begin{tabular}{cc}
      \begin{minipage}[t]{0.5\hsize}
        \centering
        \includegraphics[keepaspectratio, scale=0.5]{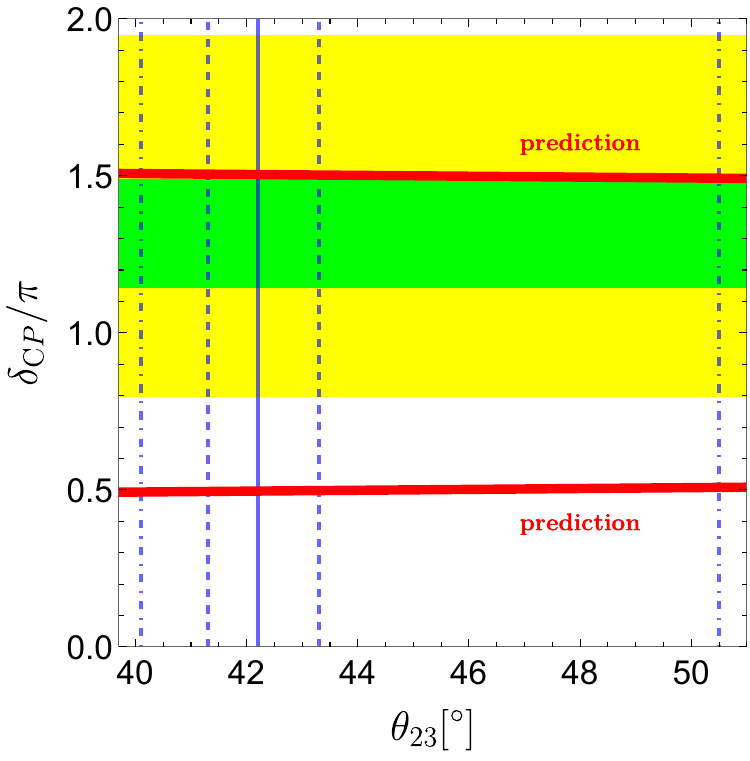}
        \subcaption{Normal ordering(NO)}
      \end{minipage} &
      \begin{minipage}[t]{0.5\hsize}
        \centering
        \includegraphics[keepaspectratio, scale=0.5]{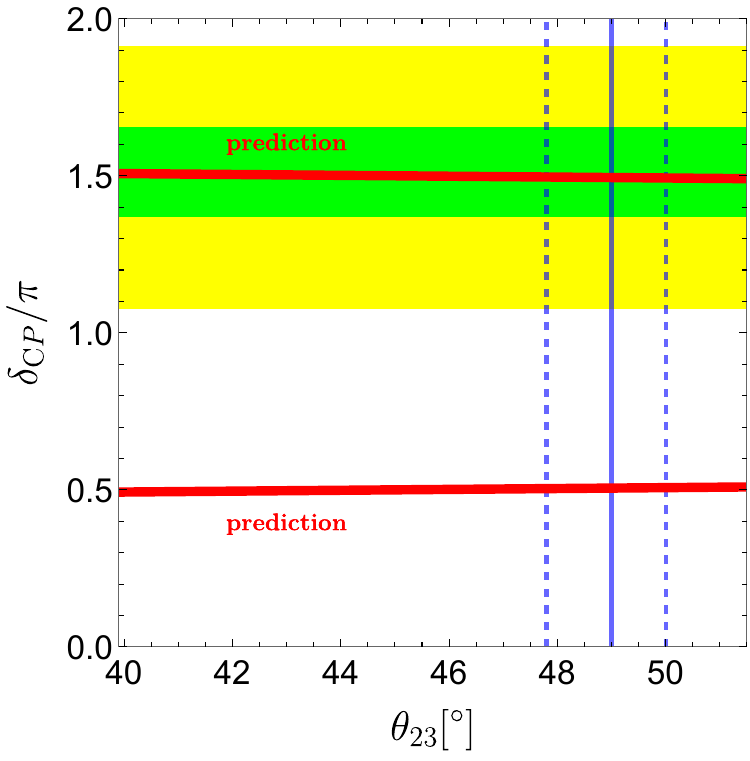}
        \subcaption{Inverted Ordering(IO)}
      \end{minipage} 
    \end{tabular}
    \caption{
    The horizontal red lines represent the predictions for the Dirac CP phase, $\delta_{\rm CP}$, as a function of $\theta_{23}$ for the ${\bf B}_3$ texture case.
    The other parameters $\theta_{12}, \theta_{13}, \Delta m_{21}^2$, and $\Delta m_{3\ell}^2$ are fixed to their best fit values.
    The green (yellow) band shows experimentally favored regions of $\delta_{\rm CP}$ at the $1\sigma$ $(3\sigma)$ level. 
    The plot range of the $x$-axis is defined by the $3\sigma$ allowed range of $\theta_{23}$. 
    The vertical blue solid line corresponds to best fit value of $\theta_{23}$, and 
    the regions between the blue dashed (dot-dashed) lines are allowed at the $1\sigma$ ($2\sigma$) level.}
    \label{fig:b3_d-theta23}
\end{figure}
%%%%%%%%%%%%%%%%%%%%%%%%%%%%%%%%%%%%%%%%%%%%%

%%%%%%%%%%%%%%%%%%% Figure %%%%%%%%%%%%%%%%%%%
\begin{figure}[tb]
    \begin{tabular}{cc}
      \begin{minipage}[t]{0.5\hsize}
        \centering
        \includegraphics[keepaspectratio, scale=0.5]{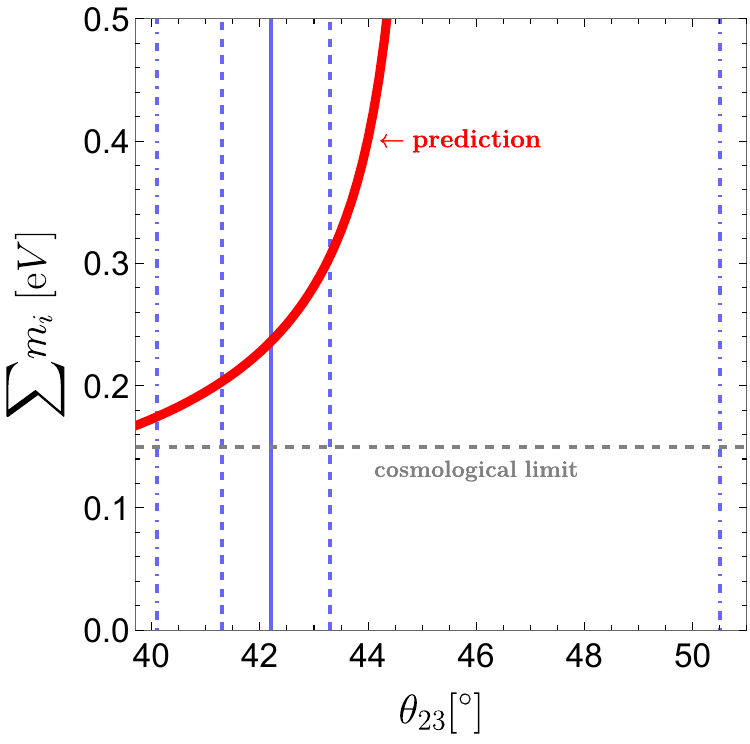}
        \subcaption{Normal ordering(NO)}
      \end{minipage} &
      \begin{minipage}[t]{0.5\hsize}
        \centering
        \includegraphics[keepaspectratio, scale=0.5]{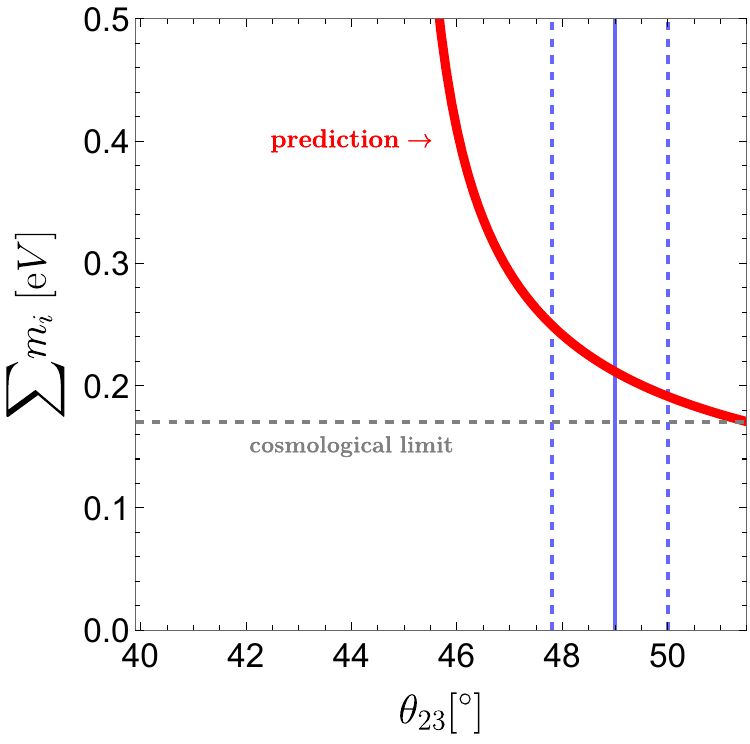}
        \subcaption{Inverted Ordering(IO)}
      \end{minipage} 
    \end{tabular}
    \caption{
    The red line represents the prediction for the sum of the light neutrino masses, $\sum m_i$, as a function of $\theta_{23}$ for the ${\bf B}_3$ texture case. 
    The other parameters $\theta_{12}, \theta_{13}, \Delta m_{21}^2$, and $\Delta m_{3\ell}^2$ are fixed to their best fit values. 
    The horizontal gray dashed line represents the 95\,\% C.L. upper bound on $\sum m_i$ from the cosmological observations: $\sum m_i \leq 0.15\,\eV$ for the NO and $\sum m_i \leq 0.17\,\eV$ for the IO~\cite{RoyChoudhury:2019hls}. 
    The vertical blue lines are the same as those in figure~\ref{fig:b3_d-theta23}.
    }
    \label{fig:b3_msum}
\end{figure}
%%%%%%%%%%%%%%%%%%%%%%%%%%%%%%%%%%%%%%%%%%%%%

In the left (right) panel of figure~\ref{fig:b3_d-theta23}, 
the horizontal red lines show non-trivial correlations between $\theta_{23}$ and $\delta_{\rm CP}$ for the ${\bf B}_3$ texture with the NO (IO). 
The parameters, $\theta_{12}, \theta_{13}, \Delta m_{21}^2$, and $\Delta m_{3\ell}^2$, are fixed to their best fit values (see table \ref{tab:NuFIT5_2}). 
The green (yellow) band shows the $1\sigma$ $(3\sigma)$ allowed range of $\delta_{\rm CP}$. 
The plot range of the $x$-axis is defined by the $3\sigma$ allowed range of $\theta_{23}$. 
The vertical blue solid line corresponds to the best fit value of $\theta_{23}$, and 
the regions between the blue dashed (dot-dashed) lines are allowed at the $1\sigma$ ($2\sigma$) level. 
It is found that the type $\mathbf{2}_{+1}$ predicts $\delta_{\rm CP} \simeq \pm \pi/2$, regardless of the neutrino mass ordering.

Similarly, we show in figure~\ref{fig:b3_msum} the prediction for the sum of the light neutrino masses $\sum m_i$ as a function of $\theta_{23}$. 
The red curve corresponds to the prediction with $\theta_{12}, \theta_{13}, \Delta m_{21}^2$, and $\Delta m_{3\ell}^2$ fixed to their best fit values. 
The horizontal gray dashed line represents the 95\,\% confidence-level (C.L.) upper bound from the cosmological observations, $\sum m_i \leq 0.15 \,(0.17)\,\eV$ for NO (IO)~\cite{RoyChoudhury:2019hls}. 
The blue lines are the same as those in figure~\ref{fig:b3_d-theta23}. 
Considering the cosmological bound on $\sum m_i$, the NO case is excluded within the $3\sigma$ range of $\theta_{23}$. 
On the other hand, the IO case marginally survives at the edge of the $3\sigma$ range, i.e., $\theta_{23} \simeq 51.5^\circ$. 
The revival of the IO case is mainly due to the shift of the allowed range of $\theta_{23}$ and the relaxation 
of the cosmological bound on $\sum m_i$. 
The predictions for the neutrino parameters at $\theta_{23} = 51.5^\circ$ are summarized in table~\ref{tab:prediction_b3}.

%%%%%%%%%%%%%%%%%%% Table %%%%%%%%%%%%%%%%%%%
\begin{table}[tb]
\centering
\begin{tabular}{|c||c|c|c|c|c|c|} 
\hline
&
$m_1$ [eV]&$m_2$ [eV]&$m_3$ [eV]&$\alpha_2/\pi$&$\alpha_3/\pi$&$\delta_{\rm CP}/\pi$
\\ \hline \hline
${\bf B}_3$ texture (IO)&$0.064$&$0.065$&$0.041$&$-0.05$&$0.96$&$1.49$
\\ \hline
\end{tabular}
\caption{
The predictions for the neutrino masses, Majorana CP phases, and the Dirac CP phase in the case of the ${\bf B}_3$ texture structure with the IO. 
The four neutrino oscillation parameters, $\theta_{12}, \theta_{13}, \Delta m_{21}^2$, and $\Delta m_{3\ell}^2$, are fixed to their best fit values. 
The mixing angle $\theta_{23}$ is fixed to $\theta_{23} = 51.5^\circ$, which is allowed at the edge of the 3$\sigma$ range. 
}
\label{tab:prediction_b3}
\end{table}
%%%%%%%%%%%%%%%%%%%%%%%%%%%%%%%%%%%%%%%%%%%%%

%%%%%%%%%%%%%%%%%%%%%%%%%%%%%%%%%%%%%%%%%%%%%%%%%%%%%%%%%%%%%%%%%%%%%%%%%%%%%%%%%%%%%%%%%%%%%%%%%%%%%%%%%%
\subsection{\texorpdfstring{${\bf B}_4$ texture}{B4 texture}}

\begin{figure}[tb]
    \begin{tabular}{cc}
      \begin{minipage}[t]{0.5\hsize}
        \centering
        \includegraphics[keepaspectratio, scale=0.5]{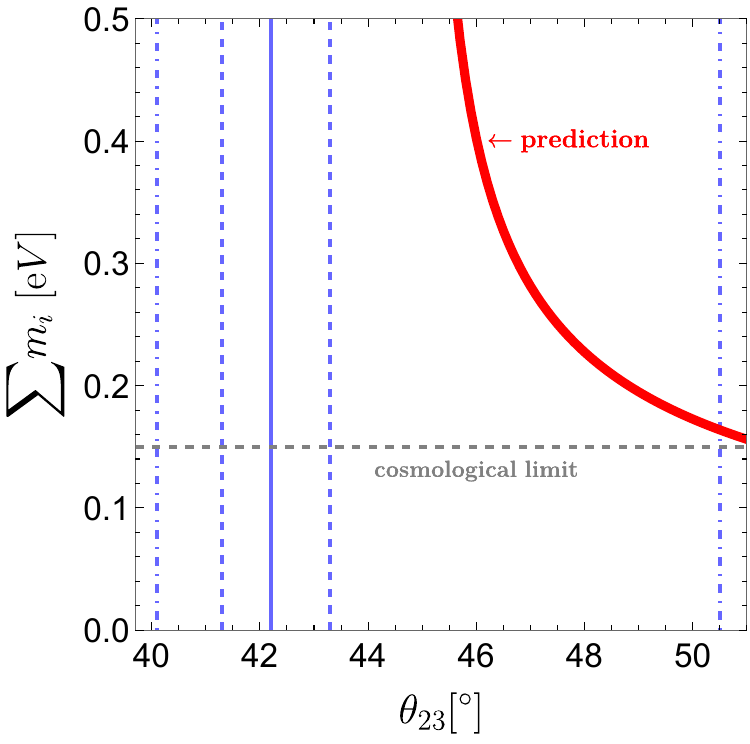}
        \subcaption{Normal ordering(NO)}
      \end{minipage} &
      \begin{minipage}[t]{0.5\hsize}
        \centering
        \includegraphics[keepaspectratio, scale=0.5]{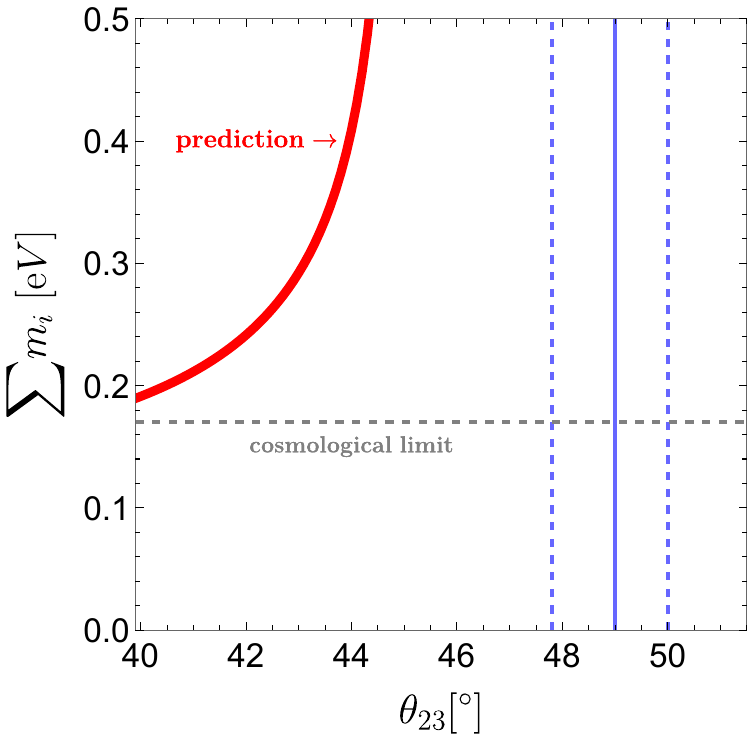}
        \subcaption{Inverted Ordering(IO)}
      \end{minipage} 
    \end{tabular}
    \caption{
    The red line shows the prediction for the sum of the light neutrino masses, $\sum m_i$, as a function of $\theta_{23}$ for the ${\bf B}_4$ texture case. 
    The parameters $\theta_{12}, \theta_{13}, \Delta m_{21}^2$, and $\Delta m_{3\ell}^2$ are fixed to their best fit values. 
    The vertical blue and horizontal gray lines are the same as those in figure~\ref{fig:b3_msum}.
    }
    \label{b4_msum}
\end{figure}

We perform the same analysis for the ${\bf B}_4$ texture case. 
The predictions for the sum of the neutrino masses are shown in figure \ref{b4_msum}. 
The vertical blue and horizontal gray lines are the same as those in figure~\ref{fig:b3_msum}. 
It is easy to see that the ${\bf B}_4$ texture is excluded for both NO and IO.

%%%%%%%%%%%%%%%%%%% Figure %%%%%%%%%%%%%%%%%%%
%%%%%%%%%%%%%%%%%%%%%%%%%%%%%%%%%%%%%%%%%%%%%

%%%%%%%%%%%%%%%%%%%%%%%%%%%%%%%%%%%%%%%%%%%%%%%%%%%%%%%%%%%%%%%%%%%%%%%%%%%%%%%%%%%%%%%%%%%%%%%%%%%%%%%%%%

\section{\texorpdfstring{New Constraints on the Gauged U(1)$_{L_\mu-L_\tau}$ Model with $\Phi_{\pm1}$}{New Constraints on the Type 2\_+- Gauged U(1)\_\{Lmu-Ltau\} Model}}
\label{sec:constraint}

We have so far focused on the structures of the neutrino mass matrix and the predictions for the neutrino parameters in the minimal gauged U$(1)_{L_\mu-L_\tau}$ models. 
The results we obtained above rely only on the two-zero matrix structures and are independent of the details of the gauge and scalar sectors. 
In this section, we study experimental constraints on the U$(1)_{L_\mu-L_\tau}$ gauge boson, with a special focus on the light gauge boson mass region ($10\,\MeV \lesssim m_{Z'}^{} \lesssim 40\,\MeV$) motivated by the muon $g-2$ anomaly. 
We will see that when the U$(1)_{L_\mu-L_\tau}$-breaking scalar is an SU(2)$_L$ non-singlet, a mixing between the $Z$ and $Z'$ bosons is induced, thereby giving rise to stringent constraints from measurements of atomic parity violation (APV) and rare meson decays. 

\subsection{\texorpdfstring{$Z-Z'$ Mixing}{Z-Z' Mixing}}

The U$(1)_{L_\mu-L_\tau}$ gauge boson $Z'$ can mix with the $Z$ boson and photon. 
We consider two types of mixings~\cite{Babu:1997st},
\begin{align}
   \label{eq:veps}
   \mathcal{L}_{\rm gauge} &= 
   -\frac{1}{4}B_{\mu\nu}B^{\mu \nu} -\frac{1}{4} Z'_{\mu\nu} Z'^{\mu \nu} + \frac{1}{2} \frac{\varepsilon}{\cos{\theta_W}} B_{\mu\nu} Z'^{\mu \nu},\\
   \label{eq:vepsZ}
    \mathcal{L}_{\varepsilon_Z^{}} &=
    \frac{m_Z^2}{2}
    \begin{pmatrix}
        Z_\mu & Z'_\mu
    \end{pmatrix}
    \begin{pmatrix}
        1 & -\varepsilon_Z^{} \\
         -\varepsilon_Z^{} & m_{Z'}^2 / m_Z^2
    \end{pmatrix}
    \begin{pmatrix}
        Z^\mu \\
        Z'^\mu
    \end{pmatrix},
\end{align}
where $B_\mu$ is the U(1)$_Y$ gauge field, 
$Z'_\mu$ is the ${\rm U}(1)_{L_\mu-L_\tau}$ gauge field,
$B_{\mu\nu} \equiv \partial_\mu B_\nu -\partial_\nu B_\mu$, 
$Z'_{\mu\nu} \equiv \partial_\mu Z'_\nu -\partial_\nu Z'_\mu$, and 
$Z_\mu \equiv \cos\theta_W \, W^3_\mu - \sin\theta_W \, B_\mu$. 
The third term in eq.~\eqref{eq:veps} is called gauge kinetic mixing which is parametrized by a dimensionless parameter $\varepsilon$.
The other mixing $\varepsilon_Z^{}$ is called mass mixing, which is generated when an SU(2)$_L$ multiplet breaks the U(1)$_{L_\mu-L_\tau}$ symmetry. 
Following the convention, the mass mixing is parametrized as 
\begin{align}
\varepsilon_Z^{} = \frac{m_{Z'}^{}}{m_Z^{}}\,\delta,
\end{align}
by a model-dependent parameter $\delta$ \cite{Davoudiasl:2012ag}. 

For $\varepsilon, \varepsilon_Z^{} \ll 1$ and $m_{Z'}^{} \ll m_Z^{}$, 
after canonically normalizing the gauge boson fields and diagonalizing the gauge boson mass matrix, the $Z'$ boson interacts with the SM fermions in the form,
\begin{align}
    \mathcal{L} \supset Z'_\mu (g_{Z'}^{} J^\mu_{L_\mu-L_\tau} + \varepsilon e J^\mu_{\rm EM} + \varepsilon_Z^{} g_Z^{} J^\mu_{\rm NC}),
    \label{eq:Z'int}
\end{align}
where $J^\mu_{L_\mu-L_\tau}$, $J^\mu_{\rm EM}$, and $J^\mu_{\rm NC}$ denote the $L_\mu-L_\tau$, electromagnetic (EM), and neutral currents, respectively, and $g_Z^{} = (g^2 +{g^\prime}^2)^{1/2}$ with $g' (g)$ being the U(1)$_Y$ (SU(2)$_L$) gauge coupling constant. 
One can see from eq.~\eqref{eq:Z'int} that $\varepsilon$ characterizes the photon--$Z'$ mixing and $\varepsilon_Z^{}$ does the $Z$--$Z'$ mixing.\footnote{For a relatively heavy $Z'$, we can include the subleading effect by taking $\delta \to \delta + \varepsilon \frac{m_{Z'}^{}}{m_Z^{}} \tan\theta_W$.} 
Note that in the parameter space of our interest, we can safely ignore the mass shifts of $Z$ and $Z'$ because of the tiny kinetic and mass mixings, i.e., $M_Z \simeq m_Z^{}$ and $M_{Z'} \simeq m_{Z'}^{}$, where $M_Z$ and $M_{Z'}$ are their physical masses. 
Hence, we simply use $m_Z^{}$ and $m_{Z'}^{}$ to denote their physical masses. 

The mass mixing $\varepsilon_Z^{}$ is generated from the $\SU(2)_L$ and $\Umt$ symmetry breaking. 
In our model, it is expressed by the masses of the $Z$ and $Z'$ bosons and the VEVs of the U(1)$_{L_\mu-L_\tau}$-breaking scalars.
On the other hand, the kinetic mixing $\varepsilon$ is a free parameter. 
In our analysis, $\varepsilon$ is assumed to be vanishing at some high-energy scale. 
In this case, a low-energy value of the kinetic mixing is determined solely by the one-loop contribution, which is given by 
\begin{align}
    \varepsilon = \frac{e g_{Z'}^{}}{12 \pi^2}\ln{\frac{m_\tau^2}{m_\mu^2}}\simeq \frac{g_{Z'}^{}}{70}, \label{eq:varepsilon}
\end{align}
where $m_{\mu (\tau)}$ is the mass of the muon (tau).

\subsection{Atomic Parity Violation}

Measurements of the APV give strong constraints on exotic parity violation.
Since the $Z'$ boson mixes with the $Z$ boson, the $Z'$ interaction with the SM fields naturally violates the parity symmetry and induces an additional contribution to the APV.

The effective Lagrangian relevant to the APV is given by 
\begin{align}
\mathcal{L}_{\rm APV} = \frac{G_F}{\sqrt{2}} \left\{ g^{eu}_{AV} (\overline{e}\gamma_\mu \gamma_5 e) (\overline{u}\gamma^\mu u) + g^{ed}_{AV} (\overline{e}\gamma_\mu \gamma_5 e)(\overline{d}\gamma^\mu d) \right\},   
\end{align}
where $G_F$ stands for the Fermi constant, and $g_{AV}^{eu}$ $(g_{AV}^{ed})$ is the four-Fermi coupling between the electron axial vector and up-quark (down-quark) vector currents.
The nuclear weak charge measured in the APV is expressed at leading order by 
\begin{align}
    Q_W^{\rm tree} = -2\left\{ N_p (2 g_{AV}^{eu} + g_{AV}^{ed}) + N_n (g_{AV}^{eu} + 2 g_{AV}^{ed}) \right\},
\end{align}
where  $N_p$ ($N_n$) is the number of protons (neutrons) in a nucleus.
In the SM, the weak charge $Q_W$ is written in the leading order by
\begin{align}
    Q_W^{\rm SM, tree} = - N_n + N_p (1 - 4 \sin^2{\theta_W}).
\end{align}

The most precise measurement of the APV is currently obtained by the $6S \to 7S$ transition in cesium~\cite{Wood:1997zq,Bennett:1999pd}. 
Recently, by applying the new data-driven determination of the neutron root-mean-square radius which is an important input to extract the atomic weak charge from the measured APV, the experimental value has been updated~\cite{Cadeddu:2021dqx};
\begin{align}
    Q^{\rm exp}_W (^{133}_{\:\;55} {\rm Cs})= -72.94(43) ,
\end{align}
while the SM value including the EW radiative corrections is given by
\begin{align}
    Q^{\rm SM}_W (^{133}_{\:\;55} {\rm Cs})= -73.23(1).
\end{align}

The $Z-Z'$ mixing modifies the atomic weak charge.
The effect from the kinetic and mass mixings can be taken into account by the following shifts in $G_F$ and $\sin^2\theta_W$~\cite{Davoudiasl:2012ag}, 
\begin{align}
    G_F &\to G_F\left( 1 + \varepsilon_Z^2 \frac{m_Z^2}{Q^2+m_{Z'}^2}  \right) =G_F\left( 1 +  \frac{m_{Z'}^2}{Q^2+m_{Z'}^2}\delta^2  \right) \label{correction_Gf}, \\
    \sin^2\theta_W &\to \sin^2\theta_W \left(1-\varepsilon \varepsilon_Z^{} \frac{\cos \theta_W}{\sin \theta_W} \frac{m_Z^2}{Q^2+m_{Z'}^2} \right), \label{correction_sin}
\end{align}
where $Q$ is the energy scale of the APV process.
For $Q^2 \ll m^2_{Z'}$, the weak charge is modified as 
\begin{align}
    Q_W 
    &\simeq Q^{\rm SM}_W (1+ \delta^2) + 4N_p \frac{\varepsilon}{\varepsilon_Z^{}}\delta^2 \cos\theta_W \sin\theta_W ,
\end{align}
where we keep the leading correction of ${\cal O}(\delta^2)$.
For the case of {}$^{133}_{\:\;55}$Cs, we have
\begin{align}
    Q_W (^{133}_{\:\;55} {\rm Cs}) \simeq Q^{\rm SM}_W (^{133}_{\:\;55} {\rm Cs}) \left\{ 1 + \delta^2 \left( 1- 1.25\ \frac{\varepsilon}{\varepsilon_Z} \right)\right\}.
\end{align}
If $Z'$ is light, we have to take propagator effects into account by replacing as $\delta^2 \to \delta^2 K_{\rm  Cs}(m_{Z^\prime})$, where $K_{\rm  Cs}(m_{Z^\prime})$ is a correction factor given in table 1 of ref.~\cite{Bouchiat:2004sp}. 
Then, the APV bound at 90\,\% C.L. is given by 
\begin{equation}
    \left|\delta^2 \left( 1- 1.25\ \frac{\varepsilon}{\varepsilon_Z^{}} \right) K_{\rm  Cs}(m_{Z'}^{})\right| \lesssim 5.7 \times 10^{-3}. 
    \label{eq:APV}
\end{equation}
For $5\,\text{MeV}\le m_{Z'}^{} \le 100\,\text{MeV}$, the correction factor takes $0.2 \le K_{\rm  Cs}\le 0.98$.
Note that for $\varepsilon / \varepsilon_Z^{} \sim 1/1.25$, the contributions from $\varepsilon$ and $\varepsilon_Z^{}$ are canceled, and the APV bound is significantly relaxed. 

\begin{comment}
\begin{table}
    \centering
    \begin{tabular}{|c||c|c|c|c|c|c|c|c|c|c|}\hline
         $m_{Z^\prime}^{}$ [Mev] & 0.1 & 0.37 & 0.5 & 1 & 2.4 & 5 & 10 & 20 & 50 & 100 \\ \hline
         $K_{\rm  Cs}(m_{Z^\prime}^{})$ & 0.025 & 0.15 & 0.20 & 0.33 & 0.5 & 0.63 & 0.74 & 0.83 & 0.93 & 0.98 \\ \hline
    \end{tabular}
    \caption{Correction factor $K_{\rm  Cs}(m_{Z^\prime}^{})$ taken from \textcolor{red}{Ref[]}.}
    \label{tab:correction_factor}
\end{table}
\end{comment}

\subsection{Flavor-changing Meson Decay}

Flavor-changing meson decay provides a good probe of a light $Z'$ boson. 
In refs.~\cite{Davoudiasl:2012ag,Davoudiasl:2014kua}, rare meson decay processes, $K \to \pi X$ and $B\to K X$ followed by $X \to e^+e^-, \mu^+\mu^-, \nu\bar{\nu}$, are analyzed for a neutral gauge boson $X$ coupled to the SM fermions only through $\varepsilon$ and $\varepsilon_Z^{}$.  
The constraints on the mixing parameters highly depend on the main decay mode. 
As for the $K^+ \to \pi^+ X$ process, for example, they obtained $\delta^2 \lesssim 10^{-4}/{\rm Br}(X \to e^+e^-)$ for $X \to e^+e^-$ and $\delta^2 \lesssim 10^{-6}/{\rm Br}(X \to \text{invisible})$ for $X \to \text{invisible}$ in the $\varepsilon \to 0$ limit. 

In our case, $Z'$ originates in the U$(1)_{L_\mu -L_\tau}$ gauge boson and couples equally to $\mu$, $\tau$, and $\nu_{\mu,\tau}$, which complicates the experimental signals in general. 
Nonetheless, in the parameter space where the muon $g-2$ discrepancy is explained, $Z'$ decays only into the neutrinos, and thus 
we consider $Z' \to \text{invisible}$ to be the main decay mode.\footnote{
The gauge boson $Z'$ can also decay into $e^+e^-$ through the kinetic and mass mixings. However, the invisible decay mode is dominant unless $e \varepsilon$ or $g_{Z}^{} \varepsilon_Z^{}$ is comparable to $g_{Z'}^{}$, which does not happen in our case.}
In this case, the $K^+ \to \pi^+ Z'$ decay provides the most stringent limit on the $Z-Z'$ mixing. 

Here, we evaluate the $K^+ \to \pi^+ Z'$ decay width, following ref.~\cite{Davoudiasl:2014kua}, and then update the experimental bound on $\delta$. 
The leading contribution to $K^+ \to \pi^+ Z'$ arises from the top-loop diagram through the mass mixing $\varepsilon_Z^{}$. 
The effect of the kinetic mixing $\varepsilon$ is 
negligible in the light $Z'$ case. 
The partial decay width is given by
\begin{align}
    \Gamma(K^+ \to \pi^+ Z') \simeq 4 \pi \frac{\sqrt{\lambda (m_K^2, m_\pi^2, m_{Z'}^2)}}{64 \pi^2 m_K^3} \times |\mathcal{M}|^2 ,
    \label{eq:decay_rate}
\end{align}
where $m_{K} \equiv m_{K^+} = 493.677\,\MeV$, $m_\pi = 139.570\,\MeV$, and 
$\lambda (a,b,c)= a^2 + b^2 + c^2 -2ab - 2ac - 2bc$ is the K\"all\'en function. 
When the SU$(2)_L$ doublet scalar $\Phi_{+1}$ is responsible for the U$(1)_{L_\mu -L_\tau}$ symmetry breaking, 
the squared decay amplitude is given by 
\begin{align}
    |\mathcal{M}|^2 = \left|\frac{g^3 V_{td}^* V_{ts} m_t^2}{128 \pi^2 m_W^3} \frac{m_Z^{}}{m_{Z'}^{}} \varepsilon_Z^{}  \left(X_1  + \frac{X_2}{\tan^2{\beta}}\right)\frac{(m_K^2 - m_\pi^2)}{2}f_+(m_{Z'}^2)\right|^2 ,
    \label{eq:amplitude}
\end{align}
where $V_{ti}\,(i=d, s)$ is an element of the Cabibbo-Kobayashi-Maskawa matrix, $m_W^{}$ is the $W$ boson mass,  
and $\tan\beta=v_1/v_2$. 
In the calculation of the decay amplitude, we use the hadronic matrix elements,
\begin{align}
    \bra{\pi^\pm(p)}\overline{s} \gamma_\mu d \ket{K^\pm(k)} &=f_+(q^2) (k+p)_\mu ,\\
    \bra{\pi^\pm(p)}\overline{s} \gamma_\mu \gamma_5 d \ket{K^\pm(k)} &= 0,
\end{align}
with $q^\mu = (k-p)^\mu$, and $f_+(0) = 1$. 
The loop functions, $X_1$ and $X_2$, are written in terms of the charged Higgs boson mass $m_{H^+}^{}$ \cite{Hall:1981bc,Frere:1981cc,PhysRevD.81.034001}: 
\begin{align}
    X_1 =&\ 2 + \frac{m_{H^+}^2}{m_{H^+}^2-m_t^2}-\frac{3 m_W^2}{m_t^2-m_W^2} + \frac{3 m_W^4(m_{H^+}^2 + m_W^2 -2m_t^2)}{(m_{H^+}^2-m_W^2)(m_t^2 - m_W^2)^2}\ln{\frac{m_t^2}{m_W^2}} \nonumber \\
    &+ \frac{m_{H^+}^2}{m_{H^+}^2-m_t^2} \left( \frac{m_{H^+}^2}{m_{H^+}^2-m_t^2} - \frac{6m_W^2}{m_{H^+}^2 - m_W^2} \ln{\frac{m_t^2}{m_{H^+}^2}} \right), \\
    X_2 =& -\frac{2m_t^2}{m_{H^+}^2-m_t^2}\left( 1+ \frac{m_{H^+}^2}{m_{H^+}^2-m_t^2} \ln{\frac{m_t^2}{m_{H^+}^2}}\right).
\end{align}
% The $X_1$ term converges to a non-zero value as $H^+$ becomes heavier, while the $X_2$ term does to zero. 
The $X_1$ term converges to a non-zero value as $m_{H^+} \to \infty$, while the $X_2$ term does to zero. 
For the full formula of the partial decay width of $K^+ \to \pi^+ Z'$ including the kinetic mixing contribution, see e.g. ref.~\cite{Davoudiasl:2014kua}. 

The branching ratio is given by
\begin{align}
    {\rm Br}(K^+ \to \pi^+ Z^\prime) = \frac{\Gamma(K^+ \to \pi^+ Z^\prime)}{\tau^{-1}_{K^+}},
    \label{eq:decay_branch}
\end{align}
where $\tau_{K^+}^{} = 1.2380 \times 10^{-8}\,\sec$ is the lifetime of the $K^+$ meson. 
For the invisibly decaying $Z'$, the current best limit on ${\rm Br}(K^+ \to \pi^+ Z^\prime)$ is reported by the NA62 experiment~\cite{NA62:2021zjw}, and we have ${\rm Br}(K^+ \to \pi^+ Z^\prime) \leq (1-6) \times 10^{-11}$ at $90 \%$ C.L. for $m_{Z'}^{} = [0, 260]\,\MeV$. 
Note that the search in ref.~\cite{NA62:2021zjw} does not give the upper limit on the branching ratio for $m_{Z'}^{} = [100, 160]\,\MeV$, because in that search all the $\pi^+$ events corresponding to that mass range are exploited to normalize the number of $K^+$ incoming.
Instead, such a mass region is relatively weakly limited by the search for invisible $\pi^0$ decays~\cite{NA62:2020pwi}.

%%%%%%%%%%%%%%%%%%% Figure %%%%%%%%%%%%%%%%%%%
\begin{figure}[tb]
    \begin{tabular}{cc}
      \begin{minipage}[t]{0.5\hsize}
        \centering
        \includegraphics[keepaspectratio, scale=0.6]{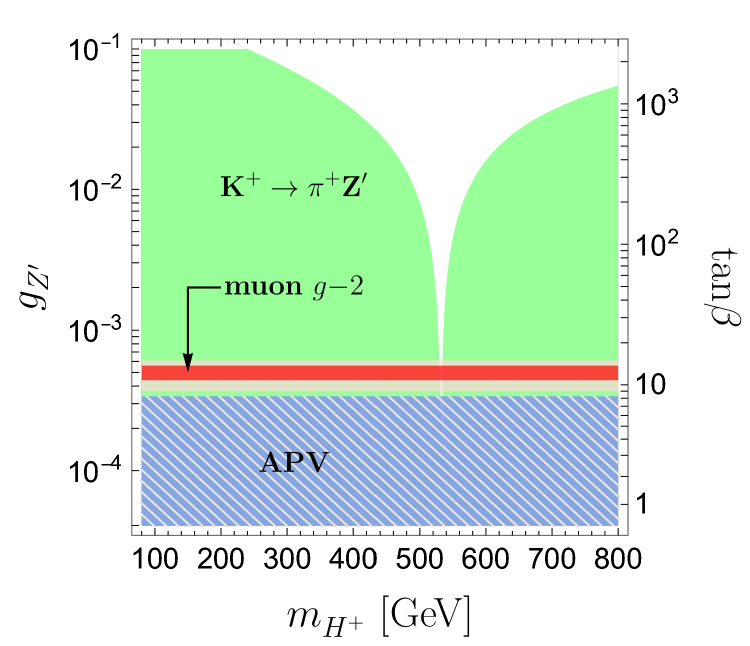}
        \subcaption{$m_{Z'}^{}=10$ MeV}
      \end{minipage} &
      \begin{minipage}[t]{0.5\hsize}
        \centering
        \includegraphics[keepaspectratio, scale=0.6]{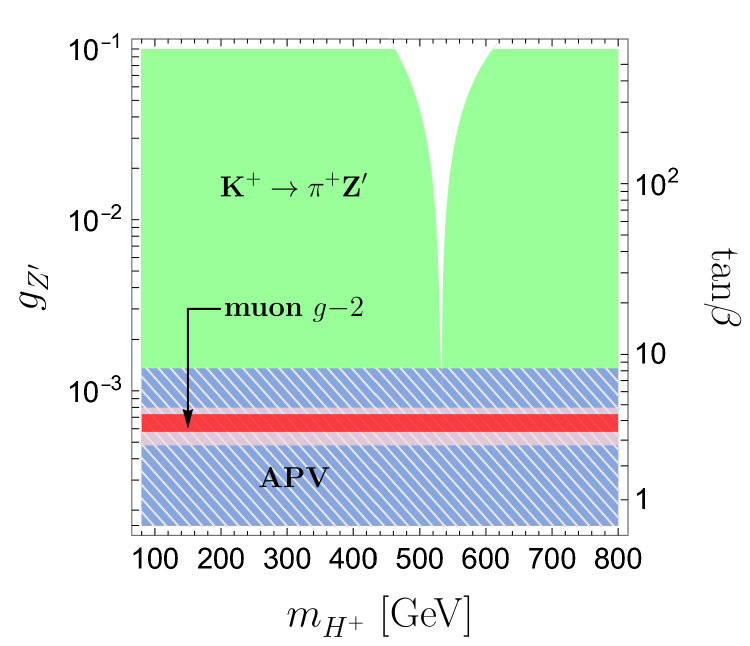}
        \subcaption{$m_{Z'}^{}=40$ MeV}
      \end{minipage} 
    \end{tabular}
    \caption{
    The experimental constraints from the APV and the flavor-changing meson decay $K^+\to\pi^+Z'$ are shown 
    for $m_{Z'}^{}=10$\,MeV (left) and $m_{Z'}^{}=40$\,MeV (right). 
    The green shaded region is excluded by $K^+\to\pi^+Z'$. 
    The blue hatched region is also constrained by the APV. 
    The (light) red band indicates the parameter region which can solve the problem of the muon $g-2$ anomaly at the $1\sigma \, (2\sigma)$ level.  
    }
    \label{FCMD_APV}
\end{figure}
%%%%%%%%%%%%%%%%%%%%%%%%%%%%%%%%%%%%%%%%%%%%%

In figure \ref{FCMD_APV}, the constraints from the Cs APV and $K^+\to\pi^+ Z'$ are shown 
% in the $(m_{H^\pm}^{},g_{Z'}^{})$ plane 
in the type $\mathbf{2}_{+1}$ model 
for $m_{Z'}^{}=10$\,MeV (left) and for $m_{Z'}^{}=40$\,MeV (right). 
We simply take $\varepsilon=0$ in the plot since the kinetic mixing has no significant impact. 
The mass of the $Z'$ boson and the mass mixing are given by 
\begin{equation}
    m_{Z'}^{} = g_{Z'} v \cos\beta , \quad
    \varepsilon_Z = \frac{m_{Z'}^{}}{m_Z^{}} \cos\beta .
\end{equation}
The blue hatched regions are excluded by the APV bound, while the green shaded regions are excluded  
by the experimental bound on $\text{Br}(K^+\to \pi^+Z')$. 
Note that for the $\text{Br}(K^+\to \pi^+Z')$ bound, a cancellation is found at $m_{H^\pm}^{}\approx 530$\,GeV, 
which corresponds to $X_1\approx 0$. 
The (light) red band indicates the favored region of the muon $g-2$ anomaly at the $1\sigma$ ($2\sigma$) level. 
Thanks to the insensitivity of the APV constraint to the charged Higgs boson mass, 
the entire parameter region favored by the muon $g-2$ anomaly is ruled out for $m_{Z'}^{}=40$\,MeV. 
On the other hand, for $m_{Z'}^{}=10$\,MeV, the region favored by the muon $g-2$ anomaly is marginally alive at $m_{H^\pm}^{}\approx 530$\,GeV.
However, we will see in the next subsection that all the parameter region favored by the muon $g-2$ anomaly 
is ruled out by constraints from the Higgs sector.

\subsection{Constraints from Higgs sector}

The heavy scalars from the second doublet $\Phi_{+1}$ modify flavor physics observables and the coupling strength of the Higgs boson to the SM particles. 
The measurements of those observables are in good agreement with the SM predictions, thereby putting strong constraints on the neutral Higgs mixing, masses of heavier scalar states as well as the VEV of the second Higgs doublet.
Here, we discuss the limits on the heavy scalars, by referring to a comprehensive study in two Higgs doublet models (2HDM) with natural flavor conservation. 

The model discussed in this paper has a similar Yukawa structure to the Type-I 2HDM. 
In that model, the strong constraints are derived from the quark flavor-changing processes \cite{Buras:2010mh, Enomoto:2015wbn, deGiorgi:2023wjh}. 
Global data for the quark flavor observables currently excludes $\tan\beta \lesssim 3$ for $m_{H^+} \simeq 300\,\GeV$, mostly due to measurements of the $B$ meson decays $B \to X_s \gamma$, $B^0_{d,s} \to \mu^+\mu^-$ and $B$ meson oscillation $\Delta M_{B_s}$~\cite{deGiorgi:2023wjh}. 
Regarding the neutral scalar mixing $\cos(\beta-\alpha)$
\footnote{The mixing angle $\alpha$ between the CP even neutral Higgs bosons is defined in appendix~\ref{app:Higgs}.}, 
the ATLAS and CMS combined limit on the Higgs coupling strength weakly constrains the Type-I 2HDM, and allows the range, $-0.35\lesssim \cos(\beta-\alpha) \lesssim 0$, at 95\,\% C.L. \cite{Bahl:2022igd}. 

The perturbative unitarity and EW precision observables also bring strong constraints to our model, given that the $\Umt$ symmetry forbids a quadratic term $(H^\dagger \Phi_{+1})$. 
In the absence of this quadratic term, the masses of the CP-even and charged Higgs bosons are determined solely by the EW VEV and scalar quartic couplings. 
Thus, it is not possible to make the additional Higgs bosons arbitrarily heavy, and their masses are restricted by the perturbative unitarity \cite{Kanemura:1993hm,Akeroyd:2000wc}.  
In appendix \ref{app:Higgs}, we provide the definition of the Higgs potential in our U(1)$_{L_\mu-L_\tau}$-gauged 2HDM as well as the notation and relations, which enable us to easily translate the unitarity bounds given in refs.~\cite{Kanemura:1993hm,Akeroyd:2000wc} 
and study the constraint from the EW oblique corrections. 
The relevant formulae for the EW oblique corrections are found in ref.~\cite{Kanemura:2011sj} with the absence of the CP odd Higgs boson.

%%%%%%%%%%%%%%%%%%% Figure %%%%%%%%%%%%%%%%%%%
\begin{figure}[tb]
        \centering
        \includegraphics[keepaspectratio, scale=0.7]{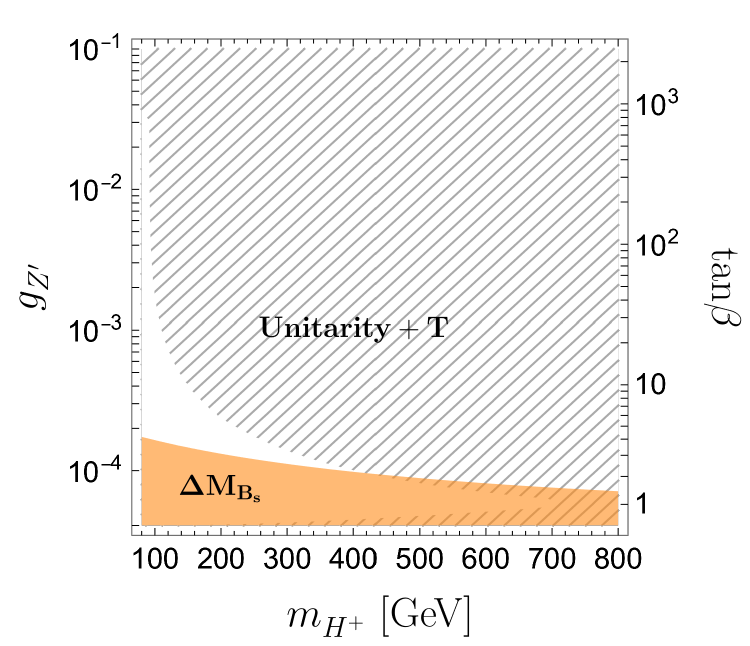}
    \caption{
    The constraints from the Higgs sector in the type $\mathbf{2}_{+1}$ model for $m_{Z'}^{}=10$ MeV. 
    The hatched region is excluded by the combination of the perturbative unitarity and EW $T$ parameter. 
    The orange shaded region is excluded by the constraint from $\Delta M_{B_s}^{}$. 
    }
    \label{B_uniT}
\end{figure}
%%%%%%%%%%%%%%%%%%%%%%%%%%%%%%%%%%%%%%%%%%%%%

In figure \ref{B_uniT}, the constraints from the Higgs sector are shown for $m_{Z'}^{}=10$\,MeV. 
Note that there is a relation between $g_{Z'}^{}$ and $\tan\beta$, 
\begin{align}
\label{massZ}
    g_{Z'}^{}=\frac{m_{Z'}^{}}{v}\sqrt{1+\tan^2\beta} .
\end{align} 
The orange shaded region is excluded by the constraint from $\Delta M_{B_s}^{}$, which gives the strongest bound among all flavor observables. 
The hatched region is excluded by the perturbative unitarity and the EW $T$ parameter. 
We scan the Higgs mixing $\cos(\beta-\alpha)$ within $-0.35\lesssim \cos(\beta-\alpha) \lesssim 0$. 
The mass ratio $m_{H^0}^{}/m_{H^\pm}^{}$ is varied from $0.1$ to $2$ in order to obtain the conservative bound. 
It is clear from figure \ref{B_uniT} that any value of $g_{Z'}^{}$ is ruled out at $m_{H^\pm}^{}\approx 530$ GeV. % by either constraint. 
Therefore, the solution of the muon $g-2$ anomaly is excluded in the type $\mathbf{2}_{+1}$ model. 

\subsection{Applications}
\label{sec:application}
Since the type $\mathbf{2}_{+1}$ model is completely ruled out, we here consider a hybrid model as an extension. 
The model features an SU(2)$_L$ singlet scalar $\sigma$ with the U(1)$_{L_\mu - L_\tau}$ charge $+1$ in addition to the SU(2)$_L$ doublet $\Phi_{\pm1}$. 
The constraints from the APV and meson decay can be relaxed by this extension. 

In the hybrid model, the $Z'$ mass and $\varepsilon_Z^{}$ are given by 
\begin{align}
m^2_{Z^\prime} 
    = g_{Z'}^2 (v_\sigma^2 + v_2^2) , \quad
\varepsilon_Z^{} = \frac{m_{Z'}^{}}{m_Z^{}} ~\text{sign}(Q_{\Phi}) \cos{\beta} \cos{\theta},
\end{align}
where $v_\sigma/\sqrt2$ is the VEV of $\sigma$, $Q_\Phi=\pm 1$ for the scalar doublet $\Phi_{\pm 1}$, and 
\begin{align}
    \tan\theta = \frac{v_\sigma}{v_2}.
\end{align}
From the definition of $\delta$, we find 
\begin{align}
    \delta = \text{sign}(Q_{\Phi}) \cos{\beta} \cos{\theta} = \frac{{\text{sign}(Q_{\Phi})}}{1+\tan^2{\theta}}  \frac1{v} \frac{m_{Z'}^{}}{ g_{Z'}^{}} .
\end{align}
Thus, for a given $m_{Z'}^{}$, 
the bounds from the APV and $K^+ \to \pi^+ Z'$ are weaken by a factor $(1+\tan^2\theta)$, 
compared with the result in figure \ref{FCMD_APV}. 
For $m_{Z'}^{}=10\,(40)$\,MeV with $\tan\theta=10\,(30)$, 
we find that the muon $g-2$ favored region is limited to the mass range $m_{H^\pm}^{}=400$--$700$ $(300$--$1000$) GeV. 
If the value of $\tan\theta$ further increases, no mass bound is obtained for the charged Higgs boson.

In this hybrid model, there is a cubic term $\mu\,H^\dag\Phi\sigma$, where $\mu$ is a parameter with mass dimension one. 
Taking a large $\mu$ with a non-negligible VEV of $\sigma$, the effects of $\Phi$ tend to decouple. 
In such a situation, the constraints from the perturbative unitarity, the EW $T$ parameter, and the Higgs mixing 
in the 2HDM sector are safely avoided. 
We also assume no mixing between the singlet sector and doublet sector without conflicting other constraints. 
Therefore, the hybrid model easily provides a viable parameter space for the solution to the muon $g-2$ discrepancy, while satisfying all other experimental and theoretical constraints.

\begin{figure}[tb]
    \centering
    \includegraphics[width=0.6 \linewidth]{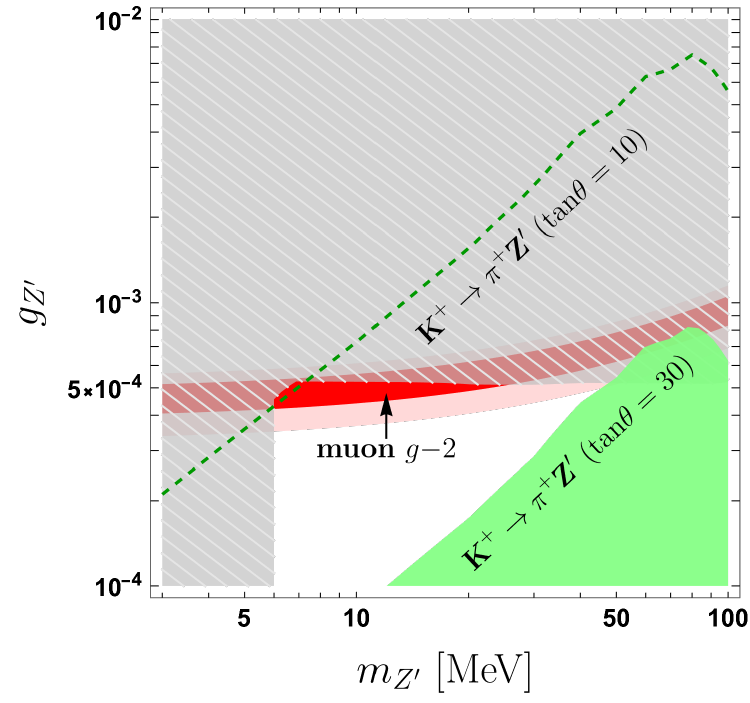}
    \caption{The constraints on the hybrid model.
    The green shaded region is excluded by the $K^+ \to \pi^+ Z^\prime$ bound with $\tan \theta = 30$. 
    The excluded region is enlarged to the region surrounded by the green dashed curve when we take $\tan\theta=10$. 
    In the (light) red region, the muon $g-2$ anomaly is explained at the $1\sigma$ $(2\sigma)$ level. 
    The gray hatched region is excluded by the NA64$\mu$ experiment~\cite{Andreev:2024sgn}, white dwarf cooling~\cite{Dreiner:2013tja,Bauer:2018onh}, and effective number of neutrinos $N_{\rm eff}$~\cite{Kamada:2018zxi,Escudero:2019gzq,Ghosh:2023ilw}.}
    \label{fig:const_s-d}
\end{figure}

In figure~\ref{fig:const_s-d}, we show the constraints from the $K^+\to\pi^+ Z^\prime$ decay as well as the well-studied U$(1)_{L_\mu - L_\tau}$ bounds. 
The latter bounds are derived by using $\varepsilon \simeq g_{Z'}^{}/70$. 
The green shaded region is excluded by the meson decay with $\tan \theta = 30$, 
where $m_{H^\pm}^{}=300$ GeV is assumed. 
When we take $\tan\theta=10$, the green shaded region is shifted to the region surrounded by the green dashed one. 
In the (light) red region, the muon $g-2$ discrepancy is explained at the $1\sigma$ $(2\sigma)$ level.
The gray hatched region is excluded by the NA64$\mu$ experiment~\cite{Andreev:2024sgn}, white dwarf cooling~\cite{Dreiner:2013tja,Bauer:2018onh}, and effective number of neutrinos $N_{\rm eff}$~\cite{Kamada:2018zxi,Escudero:2019gzq}.
Consequently, $\tan \theta$ should be larger than about $10$ to explain the muon $g-2$ discrepancy by the U(1)$_{L_\mu-L_\tau}$ gauge boson.

%%%%%%%%%%%%%%%%%%%%%%%%%%%%%%%%%%%%%%%%%%%%%%%%%%%%%%%%%%%%%%%%%%%%%%%%%%%%%%%%%%%%%%%%%%%%%%%%%%%%%%%%%%

%%%%%%%%%%%%%%%%%%%%%%%%%%%%%%%%%%%%%%%%%%%%%%%%%%%%%%%%%%%%%%%%%%%%%%%%%%%%%%%%%%%%%%%%%%%%%%%%%%%%%%
\section{Conclusion}
\label{sec:conclusion}
The minimal gauged U(1)$_{L_\mu - L_\tau}$ models, which feature a single U(1)$_{L_\mu - L_\tau}$-breaking scalar, predict various non-trivial structures for the light neutrino mass matrix. 
We have revisited the analysis of the ${\bf B}_3$ and ${\bf B}_4$ texture structures, which are realized in the models with one additional SU(2)$_L$ doublet scalar. 
Using the latest global fit, \texttt{NuFITv5.2}, of the neutrino oscillation parameters, 
we have found that only the ${\bf B}_3$ texture with the inverted mass ordering is consistent with the current neutrino oscillation data and cosmological limit on the sum of the neutrino masses. 
The allowed parameter region is very restricted and lies only at $\theta_{23} \simeq 51.5^\circ$, although the previous work concluded that this parameter space was excluded using the data sets at that time. 
We have studied the new model-dependent constraints on the U$(1)_{L_\mu - L_\tau}$ models 
with the extra SU$(2)_L$ doublet scalar by focusing on the APV and the $K^+ \to \pi^+ Z^\prime$ decay. 
The constraints from the Higgs sector, such as the quark flavor observables, perturbative unitarity, and EW precision data, 
have also been evaluated. 
It has turned out that the model is robustly ruled out by combining these constraints if the U$(1)_{L_\mu - L_\tau}$ gauge symmetry is broken solely by the SU$(2)_L$ doublet scalar. 
When an additional SU$(2)_L$ singlet scalar is also responsible for the U$(1)_{L_\mu - L_\tau}$ symmetry breaking, the constraints from the APV, meson decay, and Higgs sector are significantly relaxed. 
We have found that the VEV ratio of the singlet scalar to the doublet scalar should be larger than about $10$ to explain the muon $g-2$ discrepancy.

%%%%%%%%%%%%%%%%%%%%%%%%%%%%%%%%%%%%%%%%%%%%%%%%%%%%%%%%%%%%%%%%%%%%%%%%%%%%%%%%%%%%%%%%%%%%%%%%%%%%%%%%%%

\section*{Acknowledgements}
This work was supported by JSPS KAKENHI Grant Numbers JP23K13097 [KA], JP22K03620 [KT] and JP22K21350 [SO], and MEXT KAKENHI Grant Number JP18H05543 [KT].
%---
S.O. also acknowledges support from a Maria Zambrano fellowshop, from the State Agency for Research of the Spanish Ministry of Science and Innovation through the ``Unit of Excellence Mar\'ia de Maeztu 2020-2023'' award to the Institute of Cosmos Sciences (CEX2019-000918-M) and from PID2019-105614GB-C21 and 2017-SGR-929 grants.
%---
C.M. acknowledges Kyushu University's Innovator Fellowship Program.
%---

\appendix
%%%%%%%%%%%%%%%%%%%%%%%%%%%%%%%%%%%%%%%%%%%%%%%%%%%%%%%%%%%%%%%%%%%%%%%%%%%%%%%%%%%%%%%%%%%%%%%%%%%%%%%%%%
\section{Analysis of C minor structure}
\label{app:Cminor}

The type $\mathbf{1}$ model contains an SU(2)$_L$ singlet scalar $\sigma$ with the U(1)$_{L_\mu-L_\tau}$ charge $+1$ and vanishing hypercharge. 
This model predicts the ${\bf C}$ minor structure for the light neutrino mass matrix~\cite{Frampton:2002yf,Xing:2002ta,Xing:2002ap,Guo:2002ei,Fritzsch:2011qv,Araki:2012ip}. 
See table~\ref{tab:model1_contents} for the U(1)$_{L_\mu-L_\tau}$ charge and the SU(2)$_L$ representation of the relevant fields. 
The Lagrangian of the leptonic sector is given by 
\begin{align}
    \mathcal{L} \supset
    & -y_e\, e^c_R (L_e \cdot H^\dagger) -y_\mu\, \mu^c_R (L_\mu \cdot H^\dagger) -y_\tau\, \tau^c_R (L_\tau \cdot H^\dagger) \nonumber \\
    & -\lambda_e\, N_e^c (L_e\cdot H) -\lambda_\mu\, N_\mu^c (L_\mu\cdot H) -\lambda_\tau\, N_\tau^c (L_\tau\cdot H) \nonumber \\
    &-\frac{1}{2}M_{ee} N^c_e N^c_e -M_{\mu \tau} N^c_\mu N^c_\tau -\lambda_{e\mu} \sigma N^c_e N^c_\mu -\lambda_{e\tau} \sigma^* N^c_e N^c_\tau +{\rm H.c.}
\end{align} 
The VEV of the additional scalar $\sigma$, 
\begin{align}
\braket{\sigma} = \frac{1}{\sqrt{2}} 
    v_\sigma,
\end{align} 
is solely responsible for breaking the U$(1)_{L_\mu - L_\tau}$ gauge symmetry.
After $H$ and $\sigma$ acquire the VEVs,   
the light neutrino masses are generated through the type-I seesaw mechanism.  
The inverse of the light neutrino mass matrix takes the structure, 
\begin{align}
    \mathcal{M}_{\nu}^{-1} &\simeq - (\mathcal{M}_{D} \mathcal{M}_{R}^{-1} \mathcal{M}_{D}^T)^{-1}=
    \begin{pmatrix}
     -\frac{2 {M}_{ee}}{\lambda_e^2 v^2} & -\frac{2 \lambda_{e \mu} \braket{\sigma}}{\lambda_e \lambda _\mu v^2} & -\frac{2 \lambda_{e\tau} \braket{\sigma}}{\lambda_e \lambda_\tau  v^2} \\
     -\frac{2 \lambda_{\mu e} \braket{\sigma}}{\lambda_e \lambda_\mu  v^2} & 0 & -\frac{2 {M}_{\mu \tau}}{\lambda_\mu  \lambda_\tau  v^2} \\
     -\frac{2 \lambda_{\tau e} \braket{\sigma}}{\lambda_e \lambda_\tau  v^2} & -\frac{2 {M}_{\mu \tau}}{\lambda_\mu  \lambda_\tau  v^2} & 0
    \end{pmatrix},
\label{eq:Mnu-C}
\end{align}
where 
\begin{align}
    \mathcal{M}_{D}
    =\frac{v_1}{\sqrt{2}}
    \begin{pmatrix}
        \lambda_e  & 0 & 0 \\
        0 & \lambda_\mu  & 0 \\
        0 & 0 & \lambda_{\tau} 
    \end{pmatrix} 
    \,,\quad
    \mathcal{M}_{R}
    =
    \begin{pmatrix}
        M_{ee} & \lambda_{e \mu} \braket{\sigma} & \lambda_{e \tau} \braket{\sigma} \\
        \lambda_{\mu e} \braket{\sigma}& 0 & M_{\mu \tau} \\
        \lambda_{\tau e} \braket{\sigma} & M_{\mu \tau} & 0
    \end{pmatrix}.
\end{align}
The mass matrix structure like eq.~(\ref{eq:Mnu-C}) is called the ${\bf C}$ minor. 
\begin{table}[tb]
    \centering
    \begin{tabular}{|c||c|c|c|c|c|}\hline
         Fields & $(L_e, L_\mu, L_\tau)$ & $(e_R^{}, \mu_R^{}, \tau_R^{})$ & $(N_e, N_\mu, N_\tau)$ & $H$ & $\sigma$ \\ \hline
         U(1)$_{L_\mu - L_\tau}$ & $(0,+1,-1)$ & $(0,+1,-1)$ & $(0,+1,-1)$ & $0$ & $1$ \\ \hline
          SU(2)$_L$ & $\bf 2$ & $\bf 1$ & $\bf 1$ & $\bf 2$ & $\bf 1$ \\ \hline
    \end{tabular}
    \caption{Field contents and their charge assignments of the type $\mathbf{1}$ model.}
    \label{tab:model1_contents}
\end{table}

\begin{figure}[tb]
        \centering
        \includegraphics[keepaspectratio, scale=0.5]{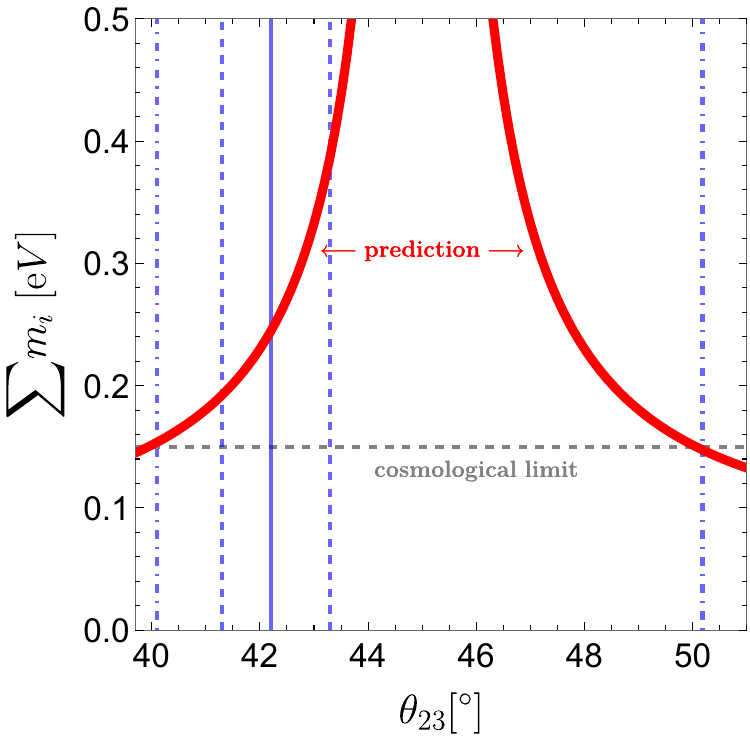}
    \caption{The prediction for the sum of the light neutrino masses, $\sum m_i$, as a function of $\theta_{23}$ for the ${\bf C}$ minor case. 
    The red curves correspond to the predictions for $\sum m_i$ when $\theta_{12}, \theta_{13}, \Delta m_{21}^2$, and $\Delta m_{3\ell}^2$ are fixed to their best fit values.
    %, and the red band is obtained by scanning only the $\theta_{12}$ value within the $1\sigma$ while fixing $\theta_{13}, \Delta m_{21}^2$, and $\Delta m_{3\ell}^2$ to their best fits. 
    The vertical blue and horizontal gray lines are the same as in figure~\ref{fig:b3_msum}.
    }
    \label{fig:msum_cminor}
\end{figure}

Following the analysis detailed in section \ref{sec:analysis}, the predictions for the neutrino parameters are obtained. 
In figure \ref{fig:msum_cminor}, 
we show the prediction for the sum of the light neutrino masses $\sum m_i$ in the case of NO.\footnote{ In the case of IO, one cannot find the real solutions of $\delta_{\rm CP}^{}$, $m_1$ and $\alpha_{2,3}$ that satisfy the two complex equations corresponding to the two zero elements of $\mathcal{M}_{\nu}^{-1}$, when the other five parameters, $\theta_{12}$, $\theta_{13}$, $\theta_{23}$, $\Delta m_{21}^2$, and $\Delta m_{3\ell}^2$, are consistent with the neutrino oscillation data.}
The red curves correspond to the predictions for $\sum m_i$ when $\theta_{12}, \theta_{13}, \Delta m_{21}^2$, and $\Delta m_{3\ell}^2$ are fixed to their best fit values.
%, and the red band is obtained by scanning only the $\theta_{12}$ value within the $1\sigma$ while fixing $\theta_{13}, \Delta m_{21}^2$, and $\Delta m_{3\ell}^2$ to their best fits.
The vertical blue and horizontal gray lines are the same as in figure~\ref{fig:b3_msum}. 
The type {\bf 1} model is allowed within the $2\sigma$ range of $\theta_{23}$.

%%%%%%%%%%%%%%%%%%%%%%%%%%%%%%%%%%%%%%%%%%%%%%%%%%%%%%%%%%%%%%%%%%%%%%%%%%%%%%%%%%%%%%%%%%%%%%%%%%%%%%%%%%
\section{Higgs potential in the type $\2_{\pm1}$ models}
\label{app:Higgs}

The Higgs potential in the type $\2_{\pm1}$ models is given by 
\begin{align}
{\mathcal V} 
= m_1^2 \, (H^\dag H) + m_2^2 \, (\Phi^\dag\Phi)
+\frac{\lambda_1}2(H^\dag H)^2
+\frac{\lambda_2}2(\Phi^\dag\Phi)^2
+\lambda_3(H^\dag H)(\Phi^\dag\Phi)
+\lambda_4(H^\dag\Phi)(\Phi^\dag H), \label{Eq:HiggsPot}
\end{align}
where $\Phi$ denotes $\Phi_{+1}$ or $\Phi_{-1}$. 
The six real parameters $m_1^2$, $m_2^2$, and $\lambda_1$--$\lambda_4$ are replaced by the physical parameters, 
the EW VEV~$v$, the neutral Higgs mixing angles~$\alpha$, 
the Higgs boson masses~$m_h^{}, m_{H^0}^{}, m_{H^+}^{}$, and the VEV ratio of the two Higgs doublet fields~$\tan\beta=\frac{v_1}{v_2}$.\footnote{The definition of $\tan\beta$ is inverted from the one in the ordinary 2HDMs.\cite{Kanemura:1993hm,Akeroyd:2000wc} } 
From the stationary conditions, we have 
\begin{align}
m_1^2 &= -\frac{v^2}2\left[\lambda_1\cos^2\beta
+(\lambda_3+\lambda_4)\sin^2\beta\right],\\
m_2^2 &= -\frac{v^2}2\left[\lambda_2\sin^2\beta
+(\lambda_3+\lambda_4)\cos^2\beta\right]. \label{Eq:sta}
\end{align}
The doublet Higgs fields are parameterized as
\begin{align}
H=\begin{pmatrix}i\,\omega_1^+\\ \frac1{\sqrt2}(v_1+h_1-i\,z_1)\end{pmatrix}, \quad
\Phi=\begin{pmatrix}i\,\omega_2^+\\ \frac1{\sqrt2}(v_2+h_2-i\,z_2)\end{pmatrix},
\end{align}
and their mixing are defined by
\begin{align}
\begin{pmatrix}h_1\\h_2\end{pmatrix}=\text{R}(\alpha)
\begin{pmatrix}H^0\\h\end{pmatrix},\quad
\begin{pmatrix}z_1\\z_2\end{pmatrix}=\text{R}(\beta)
\begin{pmatrix}z\\G\end{pmatrix},\quad
\begin{pmatrix}\omega_1^+\\\omega_2^+\end{pmatrix}=\text{R}(\beta)
\begin{pmatrix}\omega^+\\H^+\end{pmatrix},
\end{align}
where 
\begin{align}
\text{R}(\theta)=\begin{pmatrix}\cos\theta&-\sin\theta\\
\sin\theta&\cos\theta\end{pmatrix}.
\end{align}
The fields $z$ and $\omega^\pm$ are would-be Nambu-Goldstone bosons for the EW symmetry breaking, 
and $G$ is the one for the $\Umt$ symmetry breaking. 
The quartic couplings are rewritten by the physical parameters as
\begin{align}
\lambda_1 &= \frac1{v^2\sin^2\beta}\left(m_{H^0}^2\cos^2\alpha+m_h^2\sin^2\alpha\right),\\
\lambda_2 &= \frac1{v^2\cos^2\beta}\left(m_{H^0}^2\sin^2\alpha+m_h^2\cos^2\alpha\right),\\
\lambda_3 &= \frac1{v^2}\left[(m_{H^0}^2-m_h^2)\frac{\sin2\alpha}{\sin2\beta}+2m_{H^+}^2\right],\\
\lambda_4 &= \frac1{v^2}(-2m_{H^+}^2). 
\end{align}

%%%%%%%%%%%%%%%%%%%%%%%%%%%%%%%%%%%%%%%%%%%%%%%%%%%%%%%%%%%%%%%%%%%%%%%%%%%%%%%%%%%%%%%%%%%%%%%%%%%%%%%%%%

\vspace{\stretch{1}}
\newpage

\bibliography{ref}{}

\providecommand{\href}[2]{#2}\begingroup\raggedright\begin{thebibliography}{10}

\bibitem{ATLAS:2012yve}
{\scshape ATLAS} collaboration, \emph{{Observation of a new particle in the search for the Standard Model Higgs boson with the ATLAS detector at the LHC}}, \href{https://doi.org/10.1016/j.physletb.2012.08.020}{\emph{Phys. Lett. B} {\bfseries 716} (2012) 1} [\href{https://arxiv.org/abs/1207.7214}{{\ttfamily 1207.7214}}].

\bibitem{CMS:2012qbp}
{\scshape CMS} collaboration, \emph{{Observation of a New Boson at a Mass of 125 GeV with the CMS Experiment at the LHC}}, \href{https://doi.org/10.1016/j.physletb.2012.08.021}{\emph{Phys. Lett. B} {\bfseries 716} (2012) 30} [\href{https://arxiv.org/abs/1207.7235}{{\ttfamily 1207.7235}}].

\bibitem{Muong-2:2002wip}
{\scshape Muon g-2} collaboration, \emph{{Measurement of the positive muon anomalous magnetic moment to 0.7 ppm}}, \href{https://doi.org/10.1103/PhysRevLett.89.101804}{\emph{Phys. Rev. Lett.} {\bfseries 89} (2002) 101804} [\href{https://arxiv.org/abs/hep-ex/0208001}{{\ttfamily hep-ex/0208001}}].

\bibitem{Muong-2:2004fok}
{\scshape Muon g-2} collaboration, \emph{{Measurement of the negative muon anomalous magnetic moment to 0.7 ppm}}, \href{https://doi.org/10.1103/PhysRevLett.92.161802}{\emph{Phys. Rev. Lett.} {\bfseries 92} (2004) 161802} [\href{https://arxiv.org/abs/hep-ex/0401008}{{\ttfamily hep-ex/0401008}}].

\bibitem{Muong-2:2006rrc}
{\scshape Muon g-2} collaboration, \emph{{Final Report of the Muon E821 Anomalous Magnetic Moment Measurement at BNL}}, \href{https://doi.org/10.1103/PhysRevD.73.072003}{\emph{Phys. Rev. D} {\bfseries 73} (2006) 072003} [\href{https://arxiv.org/abs/hep-ex/0602035}{{\ttfamily hep-ex/0602035}}].

\bibitem{Muong-2:2021ojo}
{\scshape Muon g-2} collaboration, \emph{{Measurement of the Positive Muon Anomalous Magnetic Moment to 0.46 ppm}}, \href{https://doi.org/10.1103/PhysRevLett.126.141801}{\emph{Phys. Rev. Lett.} {\bfseries 126} (2021) 141801} [\href{https://arxiv.org/abs/2104.03281}{{\ttfamily 2104.03281}}].

\bibitem{Muong-2:2023cdq}
{\scshape Muon g-2} collaboration, \emph{{Measurement of the Positive Muon Anomalous Magnetic Moment to 0.20~ppm}}, \href{https://doi.org/10.1103/PhysRevLett.131.161802}{\emph{Phys. Rev. Lett.} {\bfseries 131} (2023) 161802} [\href{https://arxiv.org/abs/2308.06230}{{\ttfamily 2308.06230}}].

\bibitem{Aoyama:2020ynm}
T.~Aoyama et~al., \emph{{The anomalous magnetic moment of the muon in the Standard Model}}, \href{https://doi.org/10.1016/j.physrep.2020.07.006}{\emph{Phys. Rept.} {\bfseries 887} (2020) 1} [\href{https://arxiv.org/abs/2006.04822}{{\ttfamily 2006.04822}}].

\bibitem{Borsanyi:2020mff}
S.~Borsanyi et~al., \emph{{Leading hadronic contribution to the muon magnetic moment from lattice QCD}}, \href{https://doi.org/10.1038/s41586-021-03418-1}{\emph{Nature} {\bfseries 593} (2021) 51} [\href{https://arxiv.org/abs/2002.12347}{{\ttfamily 2002.12347}}].

\bibitem{Ce:2022kxy}
M.~C\`e et~al., \emph{{Window observable for the hadronic vacuum polarization contribution to the muon g-2 from lattice QCD}}, \href{https://doi.org/10.1103/PhysRevD.106.114502}{\emph{Phys. Rev. D} {\bfseries 106} (2022) 114502} [\href{https://arxiv.org/abs/2206.06582}{{\ttfamily 2206.06582}}].

\bibitem{ExtendedTwistedMass:2022jpw}
{\scshape Extended Twisted Mass} collaboration, \emph{{Lattice calculation of the short and intermediate time-distance hadronic vacuum polarization contributions to the muon magnetic moment using twisted-mass fermions}}, \href{https://doi.org/10.1103/PhysRevD.107.074506}{\emph{Phys. Rev. D} {\bfseries 107} (2023) 074506} [\href{https://arxiv.org/abs/2206.15084}{{\ttfamily 2206.15084}}].

\bibitem{CMD-3:2023alj}
{\scshape CMD-3} collaboration, \emph{{Measurement of the e+e-\textrightarrow{}\ensuremath{\pi}+\ensuremath{\pi}- cross section from threshold to 1.2~GeV with the CMD-3 detector}}, \href{https://doi.org/10.1103/PhysRevD.109.112002}{\emph{Phys. Rev. D} {\bfseries 109} (2024) 112002} [\href{https://arxiv.org/abs/2302.08834}{{\ttfamily 2302.08834}}].

\bibitem{CMD-3:2023rfe}
{\scshape CMD-3} collaboration, \emph{{Measurement of the Pion Form Factor with CMD-3 Detector and its Implication to the Hadronic Contribution to Muon (g-2)}}, \href{https://doi.org/10.1103/PhysRevLett.132.231903}{\emph{Phys. Rev. Lett.} {\bfseries 132} (2024) 231903} [\href{https://arxiv.org/abs/2309.12910}{{\ttfamily 2309.12910}}].

\bibitem{BABAR:2021cde}
{\scshape BABAR, BaBar} collaboration, \emph{{Study of the process $e^+e^-\to \pi^+\pi^-\pi^0$ using initial state radiation with BABAR}}, \href{https://doi.org/10.1103/PhysRevD.104.112003}{\emph{Phys. Rev. D} {\bfseries 104} (2021) 112003} [\href{https://arxiv.org/abs/2110.00520}{{\ttfamily 2110.00520}}].

\bibitem{Belle-II:2024msd}
{\scshape Belle-II} collaboration, \emph{{Measurement of the $e^+e^- \to \pi^+\pi^-\pi^0$ cross section in the energy range 0.62-3.50 GeV at Belle II}},  \href{https://arxiv.org/abs/2404.04915}{{\ttfamily 2404.04915}}.

\bibitem{Foot:1990mn}
R.~Foot, \emph{{New Physics From Electric Charge Quantization?}}, \href{https://doi.org/10.1142/S0217732391000543}{\emph{Mod. Phys. Lett. A} {\bfseries 6} (1991) 527}.

\bibitem{He:1990pn}
X.G.~He, G.C.~Joshi, H.~Lew and R.R.~Volkas, \emph{{NEW Z-prime PHENOMENOLOGY}}, \href{https://doi.org/10.1103/PhysRevD.43.R22}{\emph{Phys. Rev. D} {\bfseries 43} (1991) 22}.

\bibitem{He:1991qd}
X.-G.~He, G.C.~Joshi, H.~Lew and R.R.~Volkas, \emph{{Simplest Z-prime model}}, \href{https://doi.org/10.1103/PhysRevD.44.2118}{\emph{Phys. Rev. D} {\bfseries 44} (1991) 2118}.

\bibitem{Foot:1994vd}
R.~Foot, X.G.~He, H.~Lew and R.R.~Volkas, \emph{{Model for a light Z-prime boson}}, \href{https://doi.org/10.1103/PhysRevD.50.4571}{\emph{Phys. Rev. D} {\bfseries 50} (1994) 4571} [\href{https://arxiv.org/abs/hep-ph/9401250}{{\ttfamily hep-ph/9401250}}].

\bibitem{Baek:2001kca}
S.~Baek, N.G.~Deshpande, X.G.~He and P.~Ko, \emph{{Muon anomalous g-2 and gauged L(muon) - L(tau) models}}, \href{https://doi.org/10.1103/PhysRevD.64.055006}{\emph{Phys. Rev. D} {\bfseries 64} (2001) 055006} [\href{https://arxiv.org/abs/hep-ph/0104141}{{\ttfamily hep-ph/0104141}}].

\bibitem{Ma:2001md}
E.~Ma, D.P.~Roy and S.~Roy, \emph{{Gauged L(mu) - L(tau) with large muon anomalous magnetic moment and the bimaximal mixing of neutrinos}}, \href{https://doi.org/10.1016/S0370-2693(01)01428-9}{\emph{Phys. Lett. B} {\bfseries 525} (2002) 101} [\href{https://arxiv.org/abs/hep-ph/0110146}{{\ttfamily hep-ph/0110146}}].

\bibitem{Heeck:2011wj}
J.~Heeck and W.~Rodejohann, \emph{{Gauged $L_\mu - L_\tau$ Symmetry at the Electroweak Scale}}, \href{https://doi.org/10.1103/PhysRevD.84.075007}{\emph{Phys. Rev. D} {\bfseries 84} (2011) 075007} [\href{https://arxiv.org/abs/1107.5238}{{\ttfamily 1107.5238}}].

\bibitem{Harigaya:2013twa}
K.~Harigaya, T.~Igari, M.M.~Nojiri, M.~Takeuchi and K.~Tobe, \emph{{Muon g-2 and LHC phenomenology in the $L_\mu-L_\tau$ gauge symmetric model}}, \href{https://doi.org/10.1007/JHEP03(2014)105}{\emph{JHEP} {\bfseries 03} (2014) 105} [\href{https://arxiv.org/abs/1311.0870}{{\ttfamily 1311.0870}}].

\bibitem{Andreev:2024sgn}
Y.M.~Andreev et~al., \emph{{Exploration of the Muon $g-2$ and Light Dark Matter explanations in NA64 with the CERN SPS high energy muon beam}},  \href{https://arxiv.org/abs/2401.01708}{{\ttfamily 2401.01708}}.

\bibitem{Minkowski:1977sc}
P.~Minkowski, \emph{{$\mu \to e\gamma$ at a Rate of One Out of $10^{9}$ Muon Decays?}}, \href{https://doi.org/10.1016/0370-2693(77)90435-X}{\emph{Phys. Lett. B} {\bfseries 67} (1977) 421}.

\bibitem{Yanagida:1979as}
T.~Yanagida, \emph{{Horizontal gauge symmetry and masses of neutrinos}}, {\emph{Conf. Proc. C} {\bfseries 7902131} (1979) 95}.

\bibitem{Gell-Mann:1979vob}
M.~Gell-Mann, P.~Ramond and R.~Slansky, \emph{{Complex Spinors and Unified Theories}}, {\emph{Conf. Proc. C} {\bfseries 790927} (1979) 315} [\href{https://arxiv.org/abs/1306.4669}{{\ttfamily 1306.4669}}].

\bibitem{Mohapatra:1979ia}
R.N.~Mohapatra and G.~Senjanovic, \emph{{Neutrino Mass and Spontaneous Parity Nonconservation}}, \href{https://doi.org/10.1103/PhysRevLett.44.912}{\emph{Phys. Rev. Lett.} {\bfseries 44} (1980) 912}.

\bibitem{Asai:2017ryy}
K.~Asai, K.~Hamaguchi and N.~Nagata, \emph{{Predictions for the neutrino parameters in the minimal gauged U(1)$_{L_\mu-L_\tau}$ model}}, \href{https://doi.org/10.1140/epjc/s10052-017-5348-x}{\emph{Eur. Phys. J. C} {\bfseries 77} (2017) 763} [\href{https://arxiv.org/abs/1705.00419}{{\ttfamily 1705.00419}}].

\bibitem{Asai:2018ocx}
K.~Asai, K.~Hamaguchi, N.~Nagata, S.-Y.~Tseng and K.~Tsumura, \emph{{Minimal Gauged U(1)$_{L_\alpha - L_\beta}$ Models Driven into a Corner}}, \href{https://doi.org/10.1103/PhysRevD.99.055029}{\emph{Phys. Rev. D} {\bfseries 99} (2019) 055029} [\href{https://arxiv.org/abs/1811.07571}{{\ttfamily 1811.07571}}].

\bibitem{Asai:2019ciz}
K.~Asai, \emph{{Predictions for the neutrino parameters in the minimal model extended by linear combination of U(1)$_{L_e-L_\mu}$, U(1)$_{L_\mu-L_\tau}$ and U(1)$_{B-L}$ gauge symmetries}}, \href{https://doi.org/10.1140/epjc/s10052-020-7622-6}{\emph{Eur. Phys. J. C} {\bfseries 80} (2020) 76} [\href{https://arxiv.org/abs/1907.04042}{{\ttfamily 1907.04042}}].

\bibitem{Esteban:2018azc}
I.~Esteban, M.C.~Gonzalez-Garcia, A.~Hernandez-Cabezudo, M.~Maltoni and T.~Schwetz, \emph{{Global analysis of three-flavour neutrino oscillations: synergies and tensions in the determination of $\theta_{23}$, $\delta_{CP}$, and the mass ordering}}, \href{https://doi.org/10.1007/JHEP01(2019)106}{\emph{JHEP} {\bfseries 01} (2019) 106} [\href{https://arxiv.org/abs/1811.05487}{{\ttfamily 1811.05487}}].

\bibitem{Planck:2018vyg}
{\scshape Planck} collaboration, \emph{{Planck 2018 results. VI. Cosmological parameters}}, \href{https://doi.org/10.1051/0004-6361/201833910}{\emph{Astron. Astrophys.} {\bfseries 641} (2020) A6} [\href{https://arxiv.org/abs/1807.06209}{{\ttfamily 1807.06209}}].

\bibitem{nufit}
I.~Esteban, M.~Gonzalez-Garcia, M.~Maltoni, T.~Schwetz and A.~Zhou, \emph{v5.2: Three-neutrino fit based on data available in november 2022},  2022.

\bibitem{Babu:1997st}
K.S.~Babu, C.F.~Kolda and J.~March-Russell, \emph{{Implications of generalized Z - Z-prime mixing}}, \href{https://doi.org/10.1103/PhysRevD.57.6788}{\emph{Phys. Rev. D} {\bfseries 57} (1998) 6788} [\href{https://arxiv.org/abs/hep-ph/9710441}{{\ttfamily hep-ph/9710441}}].

\bibitem{Davoudiasl:2012ag}
H.~Davoudiasl, H.-S.~Lee and W.J.~Marciano, \emph{{'Dark' Z implications for Parity Violation, Rare Meson Decays, and Higgs Physics}}, \href{https://doi.org/10.1103/PhysRevD.85.115019}{\emph{Phys. Rev. D} {\bfseries 85} (2012) 115019} [\href{https://arxiv.org/abs/1203.2947}{{\ttfamily 1203.2947}}].

\bibitem{Davoudiasl:2012qa}
H.~Davoudiasl, H.-S.~Lee and W.J.~Marciano, \emph{{Muon Anomaly and Dark Parity Violation}}, \href{https://doi.org/10.1103/PhysRevLett.109.031802}{\emph{Phys. Rev. Lett.} {\bfseries 109} (2012) 031802} [\href{https://arxiv.org/abs/1205.2709}{{\ttfamily 1205.2709}}].

\bibitem{Davoudiasl:2014kua}
H.~Davoudiasl, H.-S.~Lee and W.J.~Marciano, \emph{{Muon $g-2$, rare kaon decays, and parity violation from dark bosons}}, \href{https://doi.org/10.1103/PhysRevD.89.095006}{\emph{Phys. Rev. D} {\bfseries 89} (2014) 095006} [\href{https://arxiv.org/abs/1402.3620}{{\ttfamily 1402.3620}}].

\bibitem{Frampton:2002yf}
P.H.~Frampton, S.L.~Glashow and D.~Marfatia, \emph{{Zeroes of the neutrino mass matrix}}, \href{https://doi.org/10.1016/S0370-2693(02)01817-8}{\emph{Phys. Lett. B} {\bfseries 536} (2002) 79} [\href{https://arxiv.org/abs/hep-ph/0201008}{{\ttfamily hep-ph/0201008}}].

\bibitem{Xing:2002ta}
Z.-z.~Xing, \emph{{Texture zeros and Majorana phases of the neutrino mass matrix}}, \href{https://doi.org/10.1016/S0370-2693(02)01354-0}{\emph{Phys. Lett. B} {\bfseries 530} (2002) 159} [\href{https://arxiv.org/abs/hep-ph/0201151}{{\ttfamily hep-ph/0201151}}].

\bibitem{Xing:2002ap}
Z.-z.~Xing, \emph{{A Full determination of the neutrino mass spectrum from two zero textures of the neutrino mass matrix}}, \href{https://doi.org/10.1016/S0370-2693(02)02062-2}{\emph{Phys. Lett. B} {\bfseries 539} (2002) 85} [\href{https://arxiv.org/abs/hep-ph/0205032}{{\ttfamily hep-ph/0205032}}].

\bibitem{Guo:2002ei}
W.-l.~Guo and Z.-z.~Xing, \emph{{Implications of the KamLAND measurement on the lepton flavor mixing matrix and the neutrino mass matrix}}, \href{https://doi.org/10.1103/PhysRevD.67.053002}{\emph{Phys. Rev. D} {\bfseries 67} (2003) 053002} [\href{https://arxiv.org/abs/hep-ph/0212142}{{\ttfamily hep-ph/0212142}}].

\bibitem{Fritzsch:2011qv}
H.~Fritzsch, Z.-z.~Xing and S.~Zhou, \emph{{Two-zero Textures of the Majorana Neutrino Mass Matrix and Current Experimental Tests}}, \href{https://doi.org/10.1007/JHEP09(2011)083}{\emph{JHEP} {\bfseries 09} (2011) 083} [\href{https://arxiv.org/abs/1108.4534}{{\ttfamily 1108.4534}}].

\bibitem{Araki:2012ip}
T.~Araki, J.~Heeck and J.~Kubo, \emph{{Vanishing Minors in the Neutrino Mass Matrix from Abelian Gauge Symmetries}}, \href{https://doi.org/10.1007/JHEP07(2012)083}{\emph{JHEP} {\bfseries 07} (2012) 083} [\href{https://arxiv.org/abs/1203.4951}{{\ttfamily 1203.4951}}].

\bibitem{Pontecorvo:1967fh}
B.~Pontecorvo, \emph{{Neutrino Experiments and the Problem of Conservation of Leptonic Charge}}, {\emph{Zh. Eksp. Teor. Fiz.} {\bfseries 53} (1967) 1717}.

\bibitem{Pontecorvo:1957cp}
B.~Pontecorvo, \emph{{Mesonium and anti-mesonium}}, {\emph{Sov. Phys. JETP} {\bfseries 6} (1957) 429}.

\bibitem{Pontecorvo:1957qd}
B.~Pontecorvo, \emph{{Inverse beta processes and nonconservation of lepton charge}}, {\emph{Zh. Eksp. Teor. Fiz.} {\bfseries 34} (1957) 247}.

\bibitem{Maki:1962mu}
Z.~Maki, M.~Nakagawa and S.~Sakata, \emph{{Remarks on the unified model of elementary particles}}, \href{https://doi.org/10.1143/PTP.28.870}{\emph{Prog. Theor. Phys.} {\bfseries 28} (1962) 870}.

\bibitem{Vagnozzi:2017ovm}
S.~Vagnozzi, E.~Giusarma, O.~Mena, K.~Freese, M.~Gerbino, S.~Ho et~al., \emph{{Unveiling $\nu$ secrets with cosmological data: neutrino masses and mass hierarchy}}, \href{https://doi.org/10.1103/PhysRevD.96.123503}{\emph{Phys. Rev. D} {\bfseries 96} (2017) 123503} [\href{https://arxiv.org/abs/1701.08172}{{\ttfamily 1701.08172}}].

\bibitem{Capozzi:2017ipn}
F.~Capozzi, E.~Di~Valentino, E.~Lisi, A.~Marrone, A.~Melchiorri and A.~Palazzo, \emph{{Global constraints on absolute neutrino masses and their ordering}}, \href{https://doi.org/10.1103/PhysRevD.95.096014}{\emph{Phys. Rev. D} {\bfseries 95} (2017) 096014} [\href{https://arxiv.org/abs/2003.08511}{{\ttfamily 2003.08511}}].

\bibitem{RoyChoudhury:2018gay}
S.~Roy~Choudhury and S.~Choubey, \emph{{Updated Bounds on Sum of Neutrino Masses in Various Cosmological Scenarios}}, \href{https://doi.org/10.1088/1475-7516/2018/09/017}{\emph{JCAP} {\bfseries 09} (2018) 017} [\href{https://arxiv.org/abs/1806.10832}{{\ttfamily 1806.10832}}].

\bibitem{RoyChoudhury:2019hls}
S.~Roy~Choudhury and S.~Hannestad, \emph{{Updated results on neutrino mass and mass hierarchy from cosmology with Planck 2018 likelihoods}}, \href{https://doi.org/10.1088/1475-7516/2020/07/037}{\emph{JCAP} {\bfseries 07} (2020) 037} [\href{https://arxiv.org/abs/1907.12598}{{\ttfamily 1907.12598}}].

\bibitem{Ivanov:2019hqk}
M.M.~Ivanov, M.~Simonovi\'c and M.~Zaldarriaga, \emph{{Cosmological Parameters and Neutrino Masses from the Final Planck and Full-Shape BOSS Data}}, \href{https://doi.org/10.1103/PhysRevD.101.083504}{\emph{Phys. Rev. D} {\bfseries 101} (2020) 083504} [\href{https://arxiv.org/abs/1912.08208}{{\ttfamily 1912.08208}}].

\bibitem{DES:2021wwk}
{\scshape DES} collaboration, \emph{{Dark Energy Survey Year 3 results: Cosmological constraints from galaxy clustering and weak lensing}}, \href{https://doi.org/10.1103/PhysRevD.105.023520}{\emph{Phys. Rev. D} {\bfseries 105} (2022) 023520} [\href{https://arxiv.org/abs/2105.13549}{{\ttfamily 2105.13549}}].

\bibitem{Tanseri:2022zfe}
I.~Tanseri, S.~Hagstotz, S.~Vagnozzi, E.~Giusarma and K.~Freese, \emph{{Updated neutrino mass constraints from galaxy clustering and CMB lensing-galaxy cross-correlation measurements}}, \href{https://doi.org/10.1016/j.jheap.2022.07.002}{\emph{JHEAp} {\bfseries 36} (2022) 1} [\href{https://arxiv.org/abs/2207.01913}{{\ttfamily 2207.01913}}].

\bibitem{Wood:1997zq}
C.S.~Wood, S.C.~Bennett, D.~Cho, B.P.~Masterson, J.L.~Roberts, C.E.~Tanner et~al., \emph{{Measurement of parity nonconservation and an anapole moment in cesium}}, \href{https://doi.org/10.1126/science.275.5307.1759}{\emph{Science} {\bfseries 275} (1997) 1759}.

\bibitem{Bennett:1999pd}
S.C.~Bennett and C.E.~Wieman, \emph{{Measurement of the 6S ---\ensuremath{>} 7S transition polarizability in atomic cesium and an improved test of the Standard Model}}, \href{https://doi.org/10.1103/PhysRevLett.82.2484}{\emph{Phys. Rev. Lett.} {\bfseries 82} (1999) 2484} [\href{https://arxiv.org/abs/hep-ex/9903022}{{\ttfamily hep-ex/9903022}}].

\bibitem{Cadeddu:2021dqx}
M.~Cadeddu, N.~Cargioli, F.~Dordei, C.~Giunti and E.~Picciau, \emph{{Muon and electron g-2 and proton and cesium weak charges implications on dark Zd models}}, \href{https://doi.org/10.1103/PhysRevD.104.L011701}{\emph{Phys. Rev. D} {\bfseries 104} (2021) 011701} [\href{https://arxiv.org/abs/2104.03280}{{\ttfamily 2104.03280}}].

\bibitem{Bouchiat:2004sp}
C.~Bouchiat and P.~Fayet, \emph{{Constraints on the parity-violating couplings of a new gauge boson}}, \href{https://doi.org/10.1016/j.physletb.2004.12.065}{\emph{Phys. Lett. B} {\bfseries 608} (2005) 87} [\href{https://arxiv.org/abs/hep-ph/0410260}{{\ttfamily hep-ph/0410260}}].

\bibitem{Hall:1981bc}
L.J.~Hall and M.B.~Wise, \emph{{FLAVOR CHANGING HIGGS - BOSON COUPLINGS}}, \href{https://doi.org/10.1016/0550-3213(81)90469-7}{\emph{Nucl. Phys. B} {\bfseries 187} (1981) 397}.

\bibitem{Frere:1981cc}
J.M.~Frere, J.A.M.~Vermaseren and M.B.~Gavela, \emph{{The Elusive Axion}}, \href{https://doi.org/10.1016/0370-2693(81)90686-9}{\emph{Phys. Lett. B} {\bfseries 103} (1981) 129}.

\bibitem{PhysRevD.81.034001}
M.~Freytsis, Z.~Ligeti and J.~Thaler, \emph{Constraining the axion portal with $b\ensuremath{\rightarrow}k{\ensuremath{\ell}}^{+}{\ensuremath{\ell}}^{\ensuremath{-}}$}, \href{https://doi.org/10.1103/PhysRevD.81.034001}{\emph{Phys. Rev. D} {\bfseries 81} (2010) 034001}.

\bibitem{NA62:2021zjw}
{\scshape NA62} collaboration, \emph{{Measurement of the very rare K$^{+}$\textrightarrow{}$ {\pi}^{+}\nu \overline{\nu} $ decay}}, \href{https://doi.org/10.1007/JHEP06(2021)093}{\emph{JHEP} {\bfseries 06} (2021) 093} [\href{https://arxiv.org/abs/2103.15389}{{\ttfamily 2103.15389}}].

\bibitem{NA62:2020pwi}
{\scshape NA62} collaboration, \emph{{Search for $\pi^0$ decays to invisible particles}}, \href{https://doi.org/10.1007/JHEP02(2021)201}{\emph{JHEP} {\bfseries 02} (2021) 201} [\href{https://arxiv.org/abs/2010.07644}{{\ttfamily 2010.07644}}].

\bibitem{Buras:2010mh}
A.J.~Buras, M.V.~Carlucci, S.~Gori and G.~Isidori, \emph{{Higgs-mediated FCNCs: Natural Flavour Conservation vs. Minimal Flavour Violation}}, \href{https://doi.org/10.1007/JHEP10(2010)009}{\emph{JHEP} {\bfseries 10} (2010) 009} [\href{https://arxiv.org/abs/1005.5310}{{\ttfamily 1005.5310}}].

\bibitem{Enomoto:2015wbn}
T.~Enomoto and R.~Watanabe, \emph{{Flavor constraints on the Two Higgs Doublet Models of Z$_{2}$ symmetric and aligned types}}, \href{https://doi.org/10.1007/JHEP05(2016)002}{\emph{JHEP} {\bfseries 05} (2016) 002} [\href{https://arxiv.org/abs/1511.05066}{{\ttfamily 1511.05066}}].

\bibitem{deGiorgi:2023wjh}
A.~de~Giorgi, F.~Koutroulis, L.~Merlo and S.~Pokorski, \emph{{Flavour and Higgs physics in Z2-symmetric 2HD models near the decoupling limit}}, \href{https://doi.org/10.1016/j.nuclphysb.2023.116323}{\emph{Nucl. Phys. B} {\bfseries 994} (2023) 116323} [\href{https://arxiv.org/abs/2304.10560}{{\ttfamily 2304.10560}}].

\bibitem{Bahl:2022igd}
H.~Bahl, T.~Biek\"otter, S.~Heinemeyer, C.~Li, S.~Paasch, G.~Weiglein et~al., \emph{{HiggsTools: BSM scalar phenomenology with new versions of HiggsBounds and HiggsSignals}}, \href{https://doi.org/10.1016/j.cpc.2023.108803}{\emph{Comput. Phys. Commun.} {\bfseries 291} (2023) 108803} [\href{https://arxiv.org/abs/2210.09332}{{\ttfamily 2210.09332}}].

\bibitem{Kanemura:1993hm}
S.~Kanemura, T.~Kubota and E.~Takasugi, \emph{{Lee-Quigg-Thacker bounds for Higgs boson masses in a two doublet model}}, \href{https://doi.org/10.1016/0370-2693(93)91205-2}{\emph{Phys. Lett. B} {\bfseries 313} (1993) 155} [\href{https://arxiv.org/abs/hep-ph/9303263}{{\ttfamily hep-ph/9303263}}].

\bibitem{Akeroyd:2000wc}
A.G.~Akeroyd, A.~Arhrib and E.-M.~Naimi, \emph{{Note on tree level unitarity in the general two Higgs doublet model}}, \href{https://doi.org/10.1016/S0370-2693(00)00962-X}{\emph{Phys. Lett. B} {\bfseries 490} (2000) 119} [\href{https://arxiv.org/abs/hep-ph/0006035}{{\ttfamily hep-ph/0006035}}].

\bibitem{Kanemura:2011sj}
S.~Kanemura, Y.~Okada, H.~Taniguchi and K.~Tsumura, \emph{{Indirect bounds on heavy scalar masses of the two-Higgs-doublet model in light of recent Higgs boson searches}}, \href{https://doi.org/10.1016/j.physletb.2011.09.035}{\emph{Phys. Lett. B} {\bfseries 704} (2011) 303} [\href{https://arxiv.org/abs/1108.3297}{{\ttfamily 1108.3297}}].

\bibitem{Dreiner:2013tja}
H.K.~Dreiner, J.-F.~Fortin, J.~Isern and L.~Ubaldi, \emph{{White Dwarfs constrain Dark Forces}}, \href{https://doi.org/10.1103/PhysRevD.88.043517}{\emph{Phys. Rev. D} {\bfseries 88} (2013) 043517} [\href{https://arxiv.org/abs/1303.7232}{{\ttfamily 1303.7232}}].

\bibitem{Bauer:2018onh}
M.~Bauer, P.~Foldenauer and J.~Jaeckel, \emph{{Hunting All the Hidden Photons}}, \href{https://doi.org/10.1007/JHEP07(2018)094}{\emph{JHEP} {\bfseries 07} (2018) 094} [\href{https://arxiv.org/abs/1803.05466}{{\ttfamily 1803.05466}}].

\bibitem{Kamada:2018zxi}
A.~Kamada, K.~Kaneta, K.~Yanagi and H.-B.~Yu, \emph{{Self-interacting dark matter and muon $g-2$ in a gauged U$(1)_{L_{\mu} - L_{\tau}}$ model}}, \href{https://doi.org/10.1007/JHEP06(2018)117}{\emph{JHEP} {\bfseries 06} (2018) 117} [\href{https://arxiv.org/abs/1805.00651}{{\ttfamily 1805.00651}}].

\bibitem{Escudero:2019gzq}
M.~Escudero, D.~Hooper, G.~Krnjaic and M.~Pierre, \emph{{Cosmology with A Very Light L$_{\mu}$ \ensuremath{-} L$_{\tau}$ Gauge Boson}}, \href{https://doi.org/10.1007/JHEP03(2019)071}{\emph{JHEP} {\bfseries 03} (2019) 071} [\href{https://arxiv.org/abs/1901.02010}{{\ttfamily 1901.02010}}].

\bibitem{Ghosh:2023ilw}
D.K.~Ghosh, P.~Ghosh, S.~Jeesun and R.~Srivastava, \emph{{Hubble Tension and Cosmological Imprints of $U(1)_X$ Gauge Symmetry: $U(1)_{B_3-3 L_i}$ as a case study}},  \href{https://arxiv.org/abs/2312.16304}{{\ttfamily 2312.16304}}.

\end{thebibliography}\endgroup
\bibliographystyle{JHEP}

\end{document}